\documentclass[preprint, 12pt]{elsarticle}

\usepackage{amssymb}
\usepackage{listings}

\usepackage{ifthen}
\usepackage{url}
\usepackage{wrapfig}
\usepackage{topcapt}
\usepackage{booktabs}
\usepackage{enumitem}
\usepackage{needspace}
\journal{Information and Computation}

\newboolean{hidecomments}
\setboolean{hidecomments}{false}
\ifthenelse{\boolean{hidecomments}}
{\newcommand{\nb}[2]{}}
{\newcommand{\nb}[2]{
    \fbox{\bfseries\sffamily\scriptsize#1}
    {\sf\small$\blacktriangleright$
      {#2} $\blacktriangleleft$}}}

\newcommand{\Simula}{\textsc{Simula}}
\newcommand{\code}[1]{\textsf{#1}}

\lstnewenvironment{codep}
    {\lstset{}%
      \csname lst@SetFirstLabel\endcsname}
    {\csname lst@SaveFirstLabel\endcsname}
\begin{document}

\begin{frontmatter}



\author{Andrew P. Black}
\ead{black@cs.pdx.edu}
\ead[url]{http://www.cs.pdx.edu/~black}
\address{Portland State University, Portland, Oregon, USA}

\title{Object-oriented programming: \\some history, and
challenges for the next fifty years}


\author{}

\address{}

\begin{abstract}
Object-oriented programming is inextricably linked to the pioneering work of Ole-Johan Dahl and Kristen Nygaard on the design of the \Simula{} language, which started at the Norwegian Computing Centre in the Spring of 1961.  However, object-orientation, as we think of it today\,---\,fifty years later\,---\,is the result of a complex interplay of ideas, constraints and people.  Dahl and Nygaard would certainly recognise it as their progeny,  but might also be amazed at how much it has grown up.

This article is based on a lecture given on 22$^{nd}$ August 2011, on the occasion of the scientific opening of the Ole-Johan Dahl hus at the University of Oslo.  It looks at the foundational ideas from \Simula{} that stand behind object-orientation, how those ideas have evolved to become the dominant programming paradigm, and what they have to offer as we approach the challenges of the next fifty years of informatics.

\end{abstract}


\end{frontmatter}


\section{Introduction}
\begin{wrapfigure}[16]{O}{3.2in}
   \centering
   \vspace{-1.9ex}
   \includegraphics[width=3.2in]{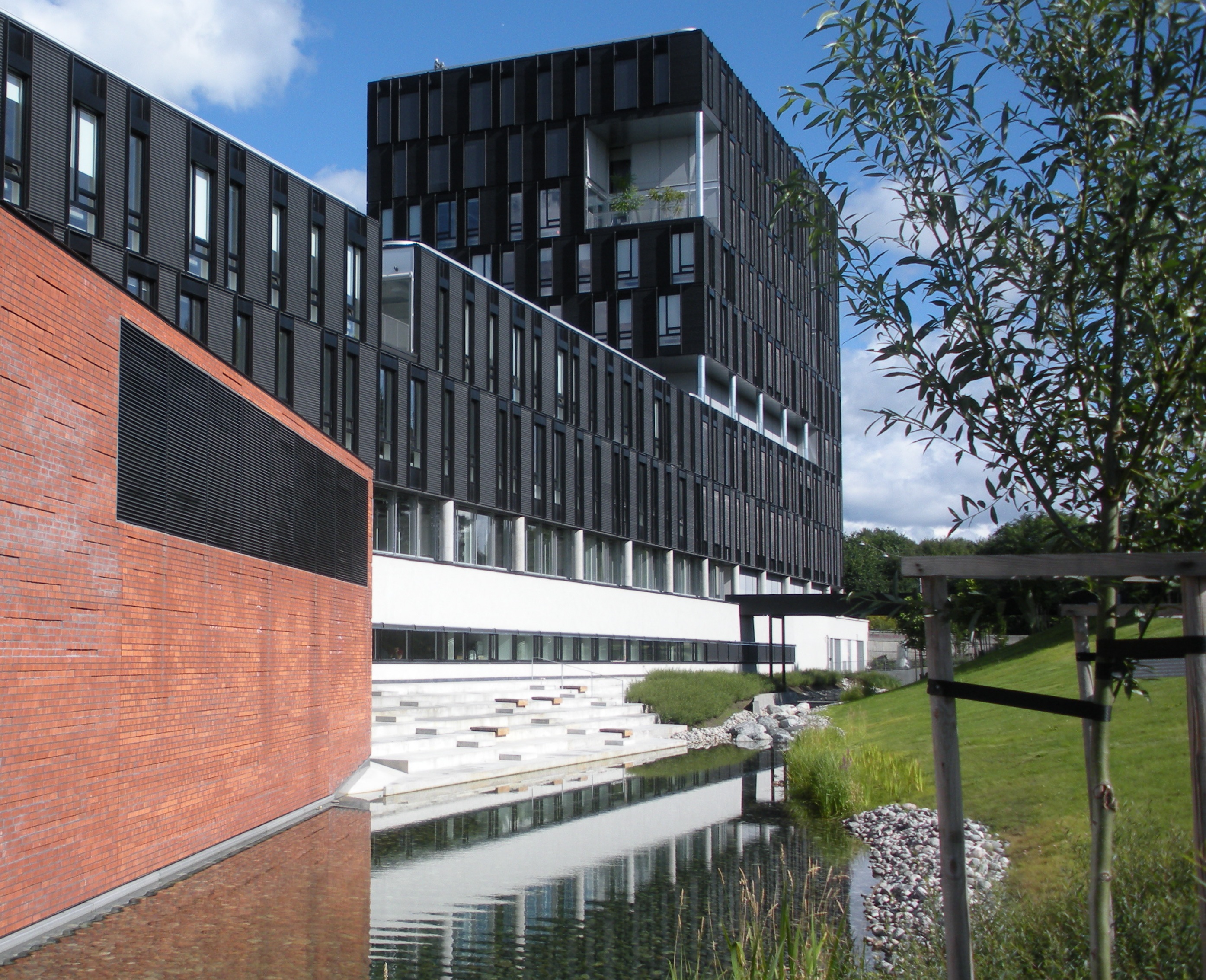} 
   \begin{minipage}{3.1in}
   \caption{The Ole-Johan Dahl hus. (Photograph \copyright{} the author)}
   \label{fig:DahlHus}
   \end{minipage}
\end{wrapfigure}

\noindent
On $22^{nd}$ August 2011, a public event was scheduled to open both the $18^{th}$ International Symposium on Fundamentals of Computation Theory and the Ole-Johan Dahl hus~\citep{fct2011}, the new building that is home to the University of Oslo's Department of Informatics, and which is shown in Figure~\ref{fig:DahlHus}. 
The morning session opened with an Introduction by Morten D\ae{}hlen, followed by two invited talks, one by myself, Andrew Black, and one by Jose Meseguer.
These talks were followed by a panel discussion on the future of object-orientation and programming languages, chaired by Arne Maus, and comprising Andrew Black, Yuri~Gurevich, Eric Jul, Jose Meseguer, and Olaf Owe.

As it happened, none of these events took place in the Dahl hus, because the beautiful lecture room that had been scheduled for the conference was put out of commission, less than half an hour before the start of the session, by an electrical fault:
the scientific opening of the Dahl hus was actually conducted in the neighbouring Kristen Nygaard building.
Thus, nine years after their deaths, Dahl and Nygaard were still able to form a symbolic partnership to solve a pressing problem. 

This article is based on the invited talk that I delivered at this event.   It is not a transcript;
I have taken the opportunity to elaborate on some themes and to pr\'{e}cises others, to add references, and to tidy up some arguments that seemed, in hindsight, a bit too ragged to set down in print.

\section{The Birth of \Simula}
\noindent
\label{intro}%
In American usage, the word ``drafted'' has many related meanings.  It can mean that you have been conscripted into military service, and it can mean that you have been given a job that is necessary, but that no one else wants to take on.

In 1948, Kristen Nygaard was drafted, in both of these senses.  He started his conscript service at the Norwegian Defence Research Establishment, where his  assignment was to carry out calculations related to the construction of Norway's first nuclear reactor~\citep{nygaar1981}.
Years later, Nygaard recalled that he had no wish to be responsible for the first nuclear accident on the continent of Europe\footnote{David Ungar, private communication.}. 

After extensive work on a traditional numerical approach, Nygaard turned to Monte Carlo simulation methods. He was made head of the ``computing office'' at the Defence Establishment, and in 1952 turned full-time to operational research.  He earned a Master of Science degree from the University of Oslo in 1956, with a thesis on probability theory entitled ``Theoretical Aspects of Monte Carlo Methods''~\citep{UnOslo2011}.  
In 1960, Nygaard moved to the Norwegian Computing Centre (NCC),  a semi-governmental research institute that had been established in 1958. 
His brief was to expand the NCC's research capabilities in computer science and operational research.  
He wrote ``Many of the civilian tasks turned out to present the same kind of methodological problems [as his earlier military work]: the necessity of using simulation, the need of concepts and a language for system description, lack of tools for generating simulation programs''~\citep{nygaar1981}.  In 1961, he started designing a simulation language as a way of attacking those problems.

In January 1962, Nygaard wrote what has become a famous letter.  It was addressed to the French operational research specialist Charles Salzmann.  
Nygaard wrote: ``The status of the Simulation Language (Monte Carlo Compiler) is that I have rather clear ideas on how to describe queueing systems, and have developed concepts which I feel allow a reasonably easy description of large classes of situations. I believe that these results have some interest even isolated from the compiler, since the presently used ways of describing such systems are not very satisfactory.  \ldots  The work on the compiler could not start before the language was fairly well developed, but this stage seems now to have been reached. The expert programmer who is interested in this part of the job will meet me tomorrow. He has been rather optimistic during our previous meetings.''

\begin{wrapfigure}[20]{O}{2.2in}
   \centering
   \vspace{-2ex}
   \includegraphics[width=2.2in]{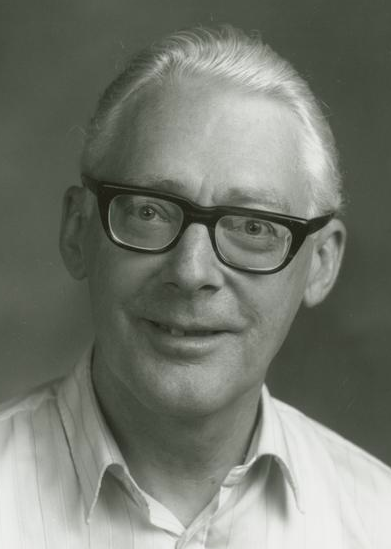} 
   \begin{minipage}{2.2in}
   \caption{Ole-Johan Dahl {\footnotesize (photograph courtesy of Department of Informatics, University of Oslo).}\label{fig:Dahl}}   
   \end{minipage}
\end{wrapfigure}

The ``expert programmer'' was of course Ole-Johan Dahl, shown in Figure~\ref{fig:Dahl}, and now widely recognised as Norway's foremost computer scientist.  Along with Nygaard, Dahl produced the initial ideas for object-oriented programming, which is now the dominant style of programming for commercial and industrial applications.  Dahl was made Commander of the Order of Saint Olav by the King of Norway in 2000, and in 2001 Dahl and Nygaard received the ACM Turing Award ``for ideas fundamental to the emergence of object-oriented programming, through their design of the programming languages \Simula{} I and \Simula{}~67.''
In 2002, Dahl and Nygaard were awarded the IEEE von Neumann medal.
Dahl died on 29$^{th}$ June 2002.

There has been some confusion about the naming of the various \Simula{} languages.
The first version of \Simula, whose preliminary design was presented in May 1963 and whose compiler was completed in January 1965, was a process description and simulation language, and was at the time named just \Simula{}.
This language had neither classes nor inheritance. 
In their paper presented at the first History of Programming Languages Conference~\citep{nygaar1981}, Nygaard and Dahl refer to this language as \Simula{} I, and I shall do the same here.
Combining their experience with this language with some important new ideas, Dahl and Nygaard went on to design a general purpose language, which was finished in 1967.  This language was at the time named \Simula~67, although there is now a Simula '87 standard, and this language is now usually referred to simply as Simula; it was \Simula~67 that
explored the ideas that were to become ``fundamental to the emergence of object-oriented programming.''  I will use the name \Simula{} when referring to the common thread of ideas that pervade both languages, and the forms with post-scripted numerals when referring to a specific language.

In this article, I will try to identify the core concepts embodied in Dahl and Nygaard's early languages, and see how these concepts have evolved in the fifty years that have passed since their invention.  I will also hazard some guesses as to how they will adapt to the future.

\section{History in Context}
\noindent
In seeking guidance for the perilous process of predicting the future, I looked at another event that also took place in 1961: the Centennial Celebration of the Massachusetts Institute of Technology.  Richard Feynman joined Sir John Cockcroft (known for splitting the atom),  Rudolf Peierls (who first conceptualised a bomb based on U-{235}) and Chen Ning Yang (who received the 1957 Nobel Prize for his work on parity laws for elementary particles) to speak on ``The Future in the Physical Sciences.''
While the other speakers contented themselves with predicting 10, or perhaps 25, years ahead, 
Feynman, who was not to receive his own Nobel Prize for another four years, decided to be really safe by predicting 1000 years ahead.
He said ``I do not think that you can read history without wondering what is the future of your own field, in a wider sense. I do not think that you can predict the future of physics alone [without] the context of the political and social world in which it lies.''
Feynman argued that in 1000 years the discovery of fundamental physical laws would have ended, but his argument was based on the social and political context in which physics must operate, rather than on any intrinsic property of physics itself~\citep{feynma1961}.

If social and political context is ultimately what determines the future of science, we must start our exploration of the \Simula{} languages by looking at the context that gave them birth.  
Nygaard's concern with modelling the phenomena associated with nuclear fission meant that \Simula{} was designed as a \emph{process description} language as well as a \emph{programming} language. ``When \Simula{} I was put to practical work it turned out that to a large extent it was used as a system description language. A common attitude among its simulation users seemed to be: sometimes actual simulation runs on the computer provided useful information.  The writing of the \Simula{} program was almost always useful, since \ldots{} it resulted in a better understanding of the system''\citep{nygaar1981}.
Notice that modelling means that the actions and interactions \emph{of the objects created by the program} model the actions and interactions of the real-world objects that they are designed to simulate.
It is not the \Simula{} code that models the real-world system, but the objects created by that code%
\footnote{More precisely, the classes of a \Simula~67 program abstract from its objects, and those objects abstract from the real world.  So the program can be regarded as a \emph{meta}-model of the real world.
I return to this topic in Section~\ref{sec:meta}.}.  
The ability to see past the code to the objects that it creates seems to have been key to the success of \Simula{}'s designers.

Because of Nygaard's concern with modelling nuclear phenomena, both of the \Simula{} languages followed Algol 60 in providing what was then called ``security'':  the languages were designed to reduce the possibility of programming errors, and so that those errors that remained could be cheaply detected at run time~\citep{HoareHints}.  A combination of compile-time and run-time checks ensured that a \Simula{} program could never give rise to machine- or implementation-dependent effects, so the behaviour of a program could be explained entirely in terms of the semantics of the programming language in which it was written~\citep{nygaar1981}.  We would call this property ``type-safety''.

This was quite different from assembly languages, and even some contemporary high-level languages such as NPL~\citep{radin1978}, where references were untyped, and the consequences of mistaking a reference to a string for a reference to a floating-point number could not be explained in terms of the language itself, but could be understood only once one knew the representations of strings and floating-point numbers chosen by the language implementer.   

\Simula{} addressed ``security'' by a combination of static and dynamic checks.  References were \emph{qualified} to refer to objects of a particular class, so the attributes of those objects\,---\,their variables and methods\,---\,were known to the compiler.  These attributes could be accessed by using a \emph{connection statement} (more commonly called an \code{inspect} statement), and in \Simula~67 by means of the now-ubiquitous ``dot'' notation.   If an attribute were present in some subclasses of the statically-known class, but not in others, then accessing it required a dynamic check, which could be requested by means of one or more \code{when} clauses in the \code{inspect} statement~\citep{dahl1967}.  This form of the \code{inspect} statement, containing multiple \code{when} clauses, is thus very similar to a type-case expression in a language like Haskell.

Dahl's view of the contributions of \Simula~67, as he explained in his r\^{o}le as discussant at the first HOPL conference~\citep{dahl1981}, was the generalisation and liberalisation of the Algol 60 block to create the class.  This gave \Simula~67:
\label{sec:SimulaSummary}
\begin{enumerate}[topsep=0ex,itemsep=0.5ex]
	\item{} record structures (class blocks with variable declarations but no statements);
	\item{} procedural data abstraction (class blocks with variable and procedure declarations, but liberated from the stack discipline so that instances of a class could outlast their callers);
	\item{} processes (class blocks that continue to execute after their callers have resumed execution);
	\item{} incremental abstraction, through \emph{prefixing} one class block with the name of another class block; \label{inh} and
	\item{} modules, obtained by prefixing an ordinary, Algol-60--style, in-line block with the name of a class block; this let the prefix block play the r\^{o}le of what Dahl called a context object\label{context}.
\end{enumerate}
Nowadays, item~\ref{inh} is called inheritance, or subclassing; I will use the terms interchangeably.
It remains one of the signature features of object-oriented programming; 
I'll return to inheritance in Section~\ref{sec:inheritance}.
In contrast, item~\ref{context}\,---\,using classes as a packaging or modularity construct\,---\,has been forgotten by mainstream languages.
Dahl wrote:
``This [item~\ref{context}] turned out to be a very interesting application, rather like collecting together sets of interrelated concepts into a larger class, and then using that larger class as a prefix to a block. The class would function as a kind of predefined context that would enable the programmer to program in a certain style directed toward a certain application area. Simulation, of course, was one of the ideas that we thought of, and we included a standard class in the language for that particular purpose''~\citep{dahl1981}.
Bracha's Newspeak language has revived this idea~\citep{bracha2010}, and the Grace language~\citep{black2012} does something very similar by using \emph{objects} as modules.

This list of what could be done with a \Simula{} class might lead one to think that the real contribution of \Simula~67{} was the class construct, and that classes, rather than the programming style that they enabled, were the real contribution of \Simula~67.
However, I don't believe that this is correct, either technically or historically.
Dahl himself wrote ``I know that \Simula{} has been criticized for perhaps having put too many things into that single basket of class.  Maybe that is correct; I'm not sure myself.  \ldots{}
It was great fun to see how easily the block concept could be remodeled and used for all these purposes. 
It is quite possible, however, that it would have been wiser to introduce a few more specialized concepts, for instance, a ``context'' concept for the contextlike classes''~\citep{dahl1981}.
I interpret these remarks as Dahl saying that the important contributions of \Simula~67 were the five individual features, and not the fact that they were all realised by a single piece of syntax.  
Using identical syntax for features that users perceive as different may be a mistake.

\section{The Origin of \Simula{}'s Core Ideas}
\noindent
Dahl was inspired to create the \Simula~67 class by visualising the \emph{runtime representation} of an Algol 60 program. 
To those with sufficient vision, objects were already in existence inside every executing Algol program\,---\,they just needed to be freed from the ``stack discipline''.
In 1972 Dahl and Hoare wrote in \textit{Hierarchical Program Structures}~\citep{dahl1972}:
\begin{quote}
\uchyph=0
In \textsc{Algol~60}, the rules of the language have been carefully designed to ensure that the lifetimes of block instances are nested, in the sense that those instances that are latest activated are the first to go out of existence. It is this feature that permits an \textsc{Algol~60} implementation to take advantage of a stack as a method of dynamic storage allocation and relinquishment. But it has the disadvantage that a program which creates a new block instance can never interact with it as an object which exists and has attributes, since it has disappeared by the time the calling program regains control. Thus the calling program can observe only the results of the actions of the procedures it calls. Consequently, the operational aspects of a block are overemphasised; and algorithms (for example, matrix multiplication) are the only concepts that can be modelled.
\end{quote} 

\begin{wrapfigure}[19]{o}{3.2in} 
   \centering
   \vspace{-1.5ex}
   \includegraphics[width=3.1in]{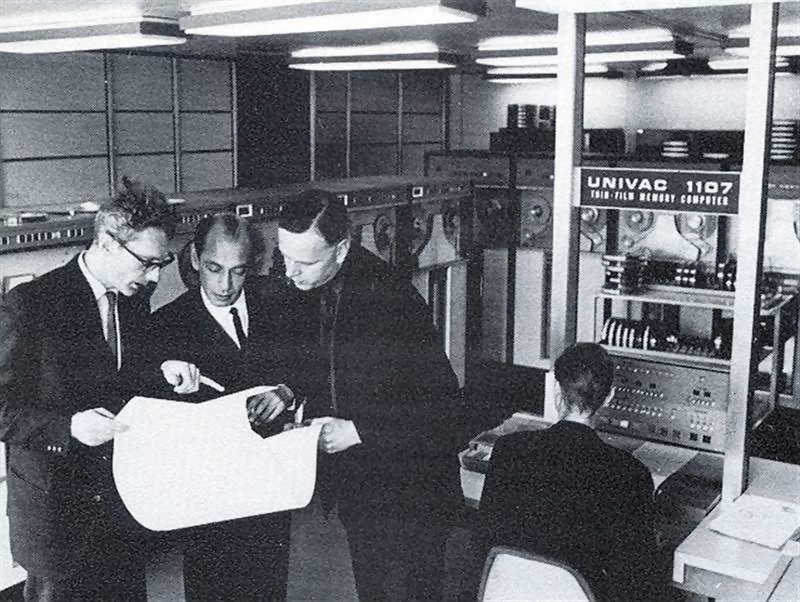} 
   \begin{minipage}{3in}
   \caption{Dahl and Nygaard during the development of \Simula{}, around 1962.  The Univac 1107 was obtained by Nygaard for the NCC at a favourable price in exchange for a contract to implement \Simula{} I.  (Photograph from the Virtual Exhibition \emph{People Behind Informatics}, \protect\url{http://cs-exhibitions.uni-klu.ac.at/})}
   \label{fig:Univac}
   \end{minipage}
\end{wrapfigure}

In \Simula~I, Dahl made two changes to the Algol 60 block: small changes, but changes with far-reaching consequences.   First, ``a block instance is permitted to outlive its calling statement, and to remain in existence for as long as the program needs to refer to it''\,\citep{dahl1972}.
Second, references to those block instances are treated as data, which gives the program a way to refer to them as independent objects.
As a consequence of these changes, a more general storage allocation mechanism than the stack is needed: a garbage collector is required to reclaim those areas of storage occupied by objects that can no longer be referenced by the running program.
Dahl and Nygaard may not have been the first to see the possibilities of generalising the Algol block, but they were the first to realise that the extra complexity of a garbage collector was a small price to pay for the wide range of concepts that could be expressed using blocks that outlive their calling creators.
It was Dahl's creation of a storage management package ``based on a two-dimensional free area list'' that made possible the \Simula{} process~\citep{nygaar1981}.
This package seems to have been completed by May 1963, the date of a preliminary presentation of the \Simula~I Language.  The \Simula{}~I compiler was completed in January 1965; a year later it was accepted by Univac, who had partially financed its development\,---\,see Figure~\ref{fig:Univac}.

\Simula~67's class concept was developed as a solution to a number of shortcomings of \Simula~I, but particularly the cumbersome nature of the \code{inspect} mechanism for accessing the local variables of a process, and the need to share properties between different simulation models.  
Hoare had previously introduced the notion of a class of record instances; one of Hoare's classes could be specialised into several subclasses, but the class and its subclasses had to be declared together, with the subclass definitions nested inside the class definition~\citep{hoare1965}.  Dahl probably became familiar with Hoare's proposals when they both lectured (Hoare on Record Handling and Dahl on Simulation Languages) at the NATO Summer school at Villard-de-Lans in 1966~\citep{genuys1968}. 
Dahl and Nygaard realised that it would be useful if a class could be declared separately from its subclasses, so that it could be specialised separately for diverse purposes.
However, Hoare's nesting of subclass declarations within class declarations precluded such a separation.  ``The solution came with the idea of class prefixing: using C as a prefix to another class, the latter would be taken to be a subclass of C inheriting all properties of C''~\citep{dahl2002}.
The semantics of prefixing was defined as concatenation, but the syntactic separation of class and subclass turned the class of \Simula~67 into a unit of reuse.   
This was so successful that the special-purpose simulation facilities of \Simula~I 
were replaced in \Simula~67 by a \textsc{simulation} class, which was defined in \Simula~67 itself.
To make use of simulation facilities in a class, all that a programmer then needed to do was to prefix the definition of their class by the name of the class \textsc{simulation}.

An important part of prefixing was the notion of \code{virtual}, although Dahl writes that it was ``a last minute extension''~\citep{dahl2002}.   
In \Simula~67, the binding of attribute identifiers is static.  
This means that the syntax of the reference, rather than the identity of the target object, determines which variable to reference or which procedure to call.   
The binding can be made dynamic by marking the attribute as \code{virtual}; 
this means that the choice of variable or procedure depends only on the target object.   Making static binding the default was, I think, a concession to efficiency; with the benefit of hindsight, it seems clear to me that it  was a mistake.  
Smalltalk made all methods virtual, and raised dynamic binding to the status of a fundamental part of object-orientation.  Even Java made dynamic binding the default, although in Java the \lstinline+final+ keyword can be used to specify static binding and to prohibit overriding.

\section{Why Inheritance Matters}
\noindent
\label{sec:inheritance}%
Inheritance, combined with dynamic binding, has become the \textit{sine qua non} of object-orientation; indeed, an influential 1987 paper by Wegner defined ``object-oriented'' as ``objects + classes + inheritance''~\citep{wegner1987}.\label{WegnerOO}
Is this inevitable?   Can one have a viable object-oriented programming language without inheritance? 
I was one who objected to Wegner's definition at the time: it seemed more reasonable for the term ``object-oriented'' to imply only the use of objects.
We had just published the first papers on Emerald~\citep{black1986c,black1987}, a language that we certainly thought of as object-oriented, even though it had no inheritance, so perhaps we were not unbiased observers.  
Wegner's definition excluded not only Emerald, but also any delegation-based language from being object-oriented, even though such languages were of growing interest.
Wegner did admit that delegation, which he defined as ``a mechanism that allows objects to delegate responsibility for performing an operation or finding a value to one or more designated `ancestors','' was actually the purer feature.

Inheritance has stood the test of time as a useful feature, and I now believe that some code-reuse mechanism \emph{like} inheritance or delegation is a very important part of any object-oriented programming system.
Since 1989, thanks to Cook, we have known that inheritance can be explained using fixpoints of generators of higher-order functions~\citep{Cook1989}.
What this means is that, in theory, functions parametrized by functions are ``as good as'' inheritance, in the sense that they can enable equivalent reuse.
Nevertheless, in practice they are not as good, because to enable reuse, the programmer has to plan ahead and make every part that could possibly change an explicit parameter.

Functional programmers call this process of explicit parametrization ``abstraction''.
It has two problems: life is uncertain, and
most people think better about the concrete than the abstract.
Let's examine these problems briefly.
We have all experienced trying to plan ahead for change.  
Inevitably, when change comes, it is not the change we anticipated.
We find that we have paid the price of generality and forward planning, but still have to modify the software to adapt to the change in requirements.  
Agile methodologies use this: they tell us to eschew generality, and instead make the software ready to embrace change by being changeable.
That is, they tell us that rather than trying to write software that is 
general enough to cope with all possible situations without need of modification,
we should write software that is specific to the current requirements, but easily modifiable 
to other possible requirements.
Inheritance aids in this process by allowing us to write software in terms of the concrete, specific actions that the application needs now, and to abstract over them only when we know\,---\,because the change request is in our hands\,---\,that the abstraction is needed.

Let's explain how this works using an example.  
Suppose that in the original program the concrete actions are represented as methods in a class.%
\footnote{The \emph{methods} of an object are the set of named operations that it can perform.
They execute in response to \emph{messages} from itself or another object; each message contains the name of a method, and the necessary arguments.  
A \emph{class} can be thought of as a template that describes the structure of a group of similarly-structured objects; in particular, it defines their methods.}
An example might be a \code{Text} class with a 
method \code{emphasise} that italicises a segment of the text.
If the requirements change so that the visual appearance of emphasis should 
controlled by a preference, then a new class can be created that inherits from \code{Text} and 
overrides the \code{emphasise} method with a new method;
this new method would use a preference object to 
determine the visual appearance of the emphasised text.
In doing so, the \code{emphasise} 
method, previously a constant, has become a parameter.
Notice that we didn't have to anticipate that this particular change would be requested, and the
original \code{Text} class didn't have to make every possible locus of change a parameter.
Even if the process of emphasising text had not been broken out into a separate method in the original code, it could have been extracted into an \code{emphasise} method using a simple, behaviour-preserving refactoring before any change to the functionality of the program was attempted.

As with so much of object-oriented programming, Dahl foresaw the value of turning methods\,---\,which he called virtual procedures\,---\,into parameters.  Indeed, virtual procedures were originally conceived, in part, as a replacement for what he felt to be the excessive power of Algol 60's call-by-name parameters.  In the discussion at the Lysebu conference following the presentation of the paper on Classes and Subclasses, Dahl is reported to have remarked~\citep{dahl1967} ``virtual quantities are in many ways similar to Algol's name parameters, but not quite as powerful.  It turns out that there is no analogy to Jensen's device.  This, I feel, is a good thing, because I hate to implement Jensen's device.  It is awful.''

The second problem with abstraction derives from the way that we think.  
Most of us can grasp new ideas most easily if we are first introduced to one or two concrete instances, and are then shown the generalisation. 
To put this another way: people learn best from examples.
So we might first solve a problem for $n = 4$, and \emph{then} make the changes necessary for $4$ to approach infinity.
There \emph{are} people who are able to first grasp abstractions directly, and then apply them: 
such people tend to become mathematicians, and sometimes programming language designers.   
Languages designed by and for such people tend to provide good support for creating abstractions.
Languages for the rest of us should support inheritance, which lets the programmer provide a concrete instance first, and then generalise from it later.

To illustrate the power of inheritance to make complex abstractions easier to understand, let's look at a case study from the functional programming literature.  In \textit{Programming Erlang}~\cite[Chapter 16]{armstr2007}, Armstrong introduces the OTP (Open Telecom Platform) generic server.  He comments: ``This is the most important section of the entire book, so read it once, read it twice, read it 100 times\,---\,just make sure the message sinks in''.
To make sure that the message about the way that the OTP server works \emph{does} sink in, Armstrong presents us with
\begin{quote}
four little servers \ldots{} each slightly different from the last.  \textsf{server1} runs some supplied code in a server, where it responds to remote requests; \textsf{server2} makes each remote request an atomic transaction;
\textsf{server3} adds hot code swapping, and 
\textsf{server4} provides both transactions and hot code swapping.
\end{quote}
Each of these ``four little servers'' is self-contained: \textsf{server4}, for example, makes no reference to any of the preceding three servers.

Why does Armstrong describe the OTP sever in this way, rather than just presenting \textsf{server4}, which is his destination?
Because he views \textsf{server4} as too complicated for the reader to understand in one go.
Something as complex as \textsf{server4} needs to be introduced step-by-step.  However, his language, lacking inheritance (and higher-order functions) does not provide a way of capturing this stepwise development.

In an effort to understand the OTP server, I coded it up in Smalltalk.  As George Forsythe is reputed to have said: ``People have said you don't understand something until you've taught it in a class. The truth is you don't understand something until you've taught it to a computer\,---\,until you've been able to program it''~\citep{archer2011}.
First I translated \textsf{server1} into Smalltalk; I called it \textsf{BasicServer}, and it had three methods and 21 lines of code.  Then I needed to test my code, so I wrote a name server plug-in for \textsf{BasicServer}, set up unit tests, and made them pass.
In the process, as Forsyth had predicted, I gained a \emph{much} clearer understanding of how Armstrong's \textsf{server1} worked.
Thus equipped, I was able to implement \textsf{TransactionServer} by subclassing \textsf{BasicServer}, and \textsf{HotSwapServer} by subclassing \textsf{TransactionServer}, bringing me to something that was equivalent to Armstrong's \textsf{server4} in two steps, each of which added just one new concern.
Then I refactored the code to increase the commonality between \code{TransactionServer} and \code{BasicServer}, a commonality that had been obscured by following Armstrong in re-writing the whole of the main server loop to implement transactions.
This refactoring added one method and one line of code. 

Once I was done, I discussed what I had learned with Phil Wadler.  
Wadler has been thinking deeply, and writing, about functional programming since the 1980s; amongst other influential articles he has authored ``The Essence of Functional Programming''~\citep{wadler1992a} and ``Comprehending Monads''~\citep{wadler1992}.
His first reaction was that the Erlang version was simpler because it could be described in straight-line code with no need for inheritance.
I pointed out that I could refactor the Smalltalk version to remove the inheritance, simply by copying down all of the methods from the superclasses into \textsf{HotSwapServer}, but that doing so would be a \emph{bad idea}.
Why?
Because the series of three classes, each building on its superclass, \emph{explained} how \textsf{HotSwapServer} worked in much the same way that Armstrong explained it in Chapter 16 of \textit{Programming Erlang}.   
This was an ``ah-ha moment'' for Phil.

To summarise:  most people understand complex ideas incrementally, by starting with a simple
concrete example, and then taking a series of generalisation steps.
A program that uses inheritance can \emph{explain} complex behaviour incrementally, 
by starting with a simple class or object, and then generalising it in a series of inheritance steps.  
The power of inheritance is that it enables us to organise our programs incrementally, that is, in a fashion that corresponds to the way that most people think.

\section{Object-Oriented Frameworks}
\noindent
In my view, one of the most significant contributions of \Simula{} 67 was the idea of the object-oriented framework.  
By the 1960s, subroutine libraries were well-known.
A subroutine, as its name suggests, is always subordinate to the main program.
The relationship is always ``don't call us, we'll call you''.
The only exception to this is when a user-written subroutine $p$ is passed as an argument to a library subroutine $q$; in this case, $q$ can indeed call $p$.
However, in the languages of the time, any such ``callback'' to $p$ could occur only during the lifetime of the original call from the main program to $q$.
An object-oriented framework allows subroutines, but also enables the reverse relationship: 
the user's program can provide one or more components that are invoked by the framework, and the framework takes on the r\^{o}le of ``main program''. 

This reversal of r\^{o}les enabled the special-purpose simulation constructs of \Simula~I  to be expressed as a framework within \Simula~67.  
The programmer of a simulation wrote code that described the behaviour of the individual objects in the system: the nuclear fuel rods, or the air masses, or the products and customers, depending on the domain being simulated. 
The simulation framework then \emph{called} those user-defined objects.
The framework was in control, but users were able to populate it with objects that achieved their diverse goals. 

Of course, for this scheme to work, the user-provided objects called by the framework must have methods for all of the requests made of them.
Providing these methods from scratch could be a lot of work, but the work can be made much more manageable by having the user-provided objects \emph{inherit} from a framework object that already has methods for the majority of these requests; all the user need do is override the inherited behaviour selectively.

One of the most common instantiations of this idea is the user-interface framework.
In most object-oriented languages, objects that are to appear on the computer's screen are sent requests by the display framework.  They must respond to enquires such as those that ask for their bounds on the screen, and to draw themselves within these bounds.  This is normally achieved by having displayable objects subclass a framework-provided class, such as Squeak's \textsf{Morph} or \textsf{java.awt.Component}. 

Dahl and Nygaard realised that these ideas could be used to provide the simulation features of \Simula~I. 
\Simula~67 is a general-purpose language; its simulation features are provided by an object-oriented framework called \textsf{\textsc{simulation}}.
\textsc{simulation} is itself implemented using a more primitive framework called  \textsc{simset}.

\section{From \Simula{} to Smalltalk}
\noindent
Smalltalk-72 was an early version of Smalltalk, designed by Alan Kay and used only within Xerox PARC.   
It is described in Kay's paper on the Early History of Smalltalk~\citep{kay1993a}.
Kay was inspired by the simplicity of LISP and the classes and objects of \Simula~67, 
and took from \Simula{} the ideas of
classes, objects and object references.

Smalltalk-72 refined and explored the idea of objects as little computers, or, as Kay puts it: ``a recursion on the notion of computer itself''.
Objects combined data with the operations on that data, in the same way that a computer combines a memory to store data with an arithmetic and logic unit to operate on the data.

Smalltalk-72 did not include inheritance, not because it was thought to be unimportant, but because \Simula's single, static inheritance seemed too limiting; 
Kay believed that inheritance could be simulated within Smalltalk using its LISP-like flexibility. 
Smalltalk-72 also dropped some other ideas from \Simula{}, notably the idea that objects were also processes, and that classes, used as prefixes to inline blocks, were also modules.
Smalltalk-76 included a Simula-like inheritance scheme that allowed superclasses to be
changed incrementally; this was kept in Smalltalk-80.
However, neither Smalltalk-76 nor Smalltalk-80 supported objects as processes or classes as modules; as a result, what we may call the ``North-American School'' of object-orientation views these features as relatively unimportant, and perhaps not really part of what it means to support objects. 

By the late 1980s, objects had attracted significant industrial interest.
Many different object-oriented languages had been developed, differing in the weight that they gave to classes, objects, inheritance, and other features.  
In 1991, Alan Snyder of Hewlett Packard  wrote an influential survey paper 
``The Essence of Objects''~\citep{snyder1991a}, which was later published in \textit{IEEE Software}~\citep{snyder1993}.

Unlike \Simula{} and Smalltalk, Snyder's is a descriptive work, not a prescriptive one.
He describes and classifies the features of contemporary object-oriented languages, including Smalltalk, C++, the Common Lisp Object System, and the Xerox Star, as well as a handful of other systems.
In Snyder's view, the essential concepts were as follows:
\begin{itemize}[itemsep=0.2ex,topsep=0.8ex]
\item{} An object embodies an abstraction.
\item{} Objects provide services.
\item{} Clients issue requests for those services.
\item{} Objects are encapsulated.
\item{} Requests identify operations.
\item{} Requests can identify objects.
\item{} New Objects can be created.
\item{} The same operation on distinct objects can have different implementations and observably different behaviour.
\item{} Objects can be classified in terms of their services (interface hierarchy).
\item{} Objects can share implementations.
	\begin{itemize}[itemsep=0.2ex,topsep=0.5ex]
	\item
	Objects can share a common implementation (multiple instances).
	\item{}
	Objects can share partial implementations (implementation inheritance or delegation).
	\end{itemize}
\end{itemize}

\begin{table}[htbp]
   \centering
   
   \topcaption{Evolution of the features of an ``object''.  \Simula's idea of procedural encapsulation  became more complete, both over time and as objects migrated across the Atlantic.  However, other features of the \Simula{} Class, such as Active Objects and Classes as Modules, have not persisted in mainstream object-oriented languages.}
\footnotesize{
   \begin{tabular}{*{3}{p{0.2\linewidth}} p{0.25\linewidth} } 

     \textbf{Feature}    & \textbf{\Simula{} 67} & \textbf{Smalltalk-80} & \textbf{Snyder (1991)} \\
     \cmidrule(lr){1-4}
	
      \parbox[t]{\linewidth}{
			\raggedright 
      				Procedural Abstraction}
      &
      \parbox[t]{\linewidth}{
			\raggedright  Attributes exposed}
      & 
      \parbox[t]{\linewidth}{
			\raggedright Attributes encapsulated}
      &
      \parbox[t]{\linewidth}{
			\raggedright  Objects characterised by services}\\ \\
      Active Objects & Yes & No & ``Associated Concept''\\[2ex]
      
      \parbox[t]{\linewidth}{
			\raggedright Dynamic object creation}
      & Yes & Yes & Yes \\ \\
      Classes & Yes & Yes & Shared implementations \\[2ex]
      Inheritance  &  Class prefixing & Subclassing 
      &
      \parbox[t]{\linewidth}{
			\raggedright  ``Shared partial \break{}implementations''} \\ \\
      \parbox[t]{\linewidth}{
			\raggedright Classes as Modules}
	& Yes & No & No \\[3ex]
      \cmidrule(lr){1-4}
   \end{tabular}   \label{tab:ObjectEvolution}
   }
\end{table}

We see that objects as processes and classes as modules are not on this list.  
Snyder does mention ``Active Objects'' as an ``associated concept'', that is, an idea ``associated with the notion of objects, but not essential to it''.  The idea that classes can serve as modules does not appear at all.

The evolution of the features that make up object-orientation is summarised in 
Table~\ref{tab:ObjectEvolution}.  Encapsulation has become more important, and the mechanisms for sharing implementation have become more diverse and refined.  
The idea that multiple objects can share an implementation, and that new kinds of objects can be defined that share parts of existing objects, have become recognised as important, 
although both kinds of sharing can be implemented either using \Simula-like classes, or through other mechanisms such as delegation.
In contrast, active objects, and classes as modules, are no longer regarded as key to object-orientation.

\section{Objects as Abstractions}
\noindent
The word ``abstraction'' is used in Informatics in many different senses.
As I mentioned previously, it is used by functional programmers to mean ``parameterisation''.
In the context of objects, I am going to use the word abstraction to capture the idea  that what matters about an object is its protocol: the set of messages that it understands, and the way that it behaves in response to those messages.  Nowadays, this is  sometimes also referred to as the object's interface.  
The key idea is that when we use an object, we focus on how it appears from the outside, and ``abstract away from'' its internal structure: more simply, that the internal structure of an object is hidden from all other objects.  That's why I said ``the set of messages that it understands'', and not ``the set of methods that it implements''.  In many languages they are the same,  but I wanted to emphasise the external rather than the internal view.
 
Abstraction, in this sense, is one of the key ideas behind objects, but it does not appear explicitly until Snyder's 1991 survey.  
\Simula{} doesn't mention abstraction specifically; it speaks instead of modelling, which captures the importance of interacting with the object, but not the information hiding aspect.
Dahl remarks~\citep{dahl2002} ``According to a comment in [the \Simula{} user's manual, 1965] it was a pity that the variable attributes of a \emph{Car} process could not be hidden away in a subblock'', but this was not possible in \Simula{} without also hiding the procedures that defined the Car's behaviour\,---\,and exposing those procedures was the whole reason for defining the Car.  (Around 1972, \Simula{} addressed this problem by adding the \code{protected} and \code{hidden} keywords to control the visibility of a name declared in a class~\citep{Krogda2010}.)

Smalltalk-80 approached abstraction by protecting the {instance variables} that contain the data representing an object: these variables are accessible only by the object itself.  The Smalltalk-80 programming environment also provided a way for the programmer to say that a method should be private, but that was merely a convention, and not enforced by the system.
Nevertheless, Dan Ingalls' sweeping 1981 \textit{Byte} article ``Design Principles behind Smalltalk''~\cite{ingall1981} refers to abstraction only indirectly.
Ingalls singles out \emph{Classification}, not abstraction,  as the key idea embodied in the Smalltalk class.  Ingalls writes: 
\label{IngallsClassification}
\begin{quote}
Classification is the objectification of \emph{ness}ness. In other words, when a human sees a chair, the experience is taken both literally as ``that very thing'' and abstractly as ``that chair-like thing''. Such abstraction results from the marvellous ability of the mind to merge ``similar'' experience, and this abstraction manifests itself as another object in the mind, the Platonic chair or chair\emph{ness}.
\end{quote}
The fact remains that Smalltalk
classes do not classify objects on the basis of their behaviour, but according to their representation.
It seems that the idea of separating the internal (concrete) and external (abstract) view of an object was yet to mature,
although it can be glimpsed elsewhere in Ingalls' article.  After Classification, his next principle is ``\emph{Polymorphism}: A program should specify only the behaviour of objects [that it uses], not their representation''.
Also, in the ``Future Work'' section, Ingalls remarks that 
\begin{quote}
message protocols have not been formalised. The organisation provides for protocols, but it is currently only a matter of style for protocols to be consistent from one class to another. This can be remedied easily by providing proper protocol objects that can be consistently shared. This will then allow formal typing of variables by protocol without losing the advantages of polymorphism.
\end{quote}
Borning and Ingalls did indeed attempt just that~\citep{bornin1982}, but it proved not to be as easy to get the details right as Ingalls had predicted.
Still, Smalltalk as defined in 1980 did succeed in abstracting away from representation details:
it never assumes that two objects that exhibit the same protocol must also have the same representation.

The related concept of \emph{data abstraction} can be traced back to the 1970s.
Landmarks in its evolution are of course Hoare's ``Proof of Correctness of Data Representations''~\citep{hoare1972}, Parnas' ``On the Criteria to be used in Decomposing Systems into Modules''~\citep{parnas1972b}, both published in 1972, and the CLU programming language, which provided explicit conversion between the concrete data representation seen by the implementation and the abstract view seen by the client~\citep{Liskov77}.
However, none of these papers pursues the idea that multiple representations of a single abstraction could co-exist.  They assume that abstraction is either a personal discipline backed by training the programmer, or a linguistic discipline backed by types.
The latter view\,---\,that abstraction comes from types\,---\,has been heavily marketed by the proponents of ML and Haskell.
It is based on the idea that types capture representation as well as interface, but that by hiding the actual type from code that lies outside the abstraction boundary, the programmer can be prohibited from taking advantage of the representation.
The mechanism used for this hiding is the notion of an \emph{existentially quantified type}.

Objects follow a different route to abstraction, which can be traced back to Alonzo Church's work on representing numbers in the lambda-calculus~\citep{church1941}.  
Church represented the number $2$ in the pure lambda-calculus by the function that \emph{did something twice}; the number $3$ by the function that did something three times, and so on.  
In 1975 John Reynolds named this route ``procedural abstraction'' and showed how it could be used to support data abstraction in a programming language, as well as contrasting it with the approach based on existential types~\citep{reynol1975}. 
Procedural abstraction is characterised by using a \emph{computational} mechanism\,---\,a representation of a function as executable code\,---\,rather than type discipline or self-discipline\,---\,to enforce abstraction boundaries.  Reynolds writes:
\begin{quote}
Procedural data structures provide a decentralised form of data abstraction.  Each part of the program which creates procedural data will specify its own form of representation, independently of the representations used elsewhere for the same kind of data, and will provide versions of the primitive operations (the components of the procedural data item) suitable for this representation.  There need be no part of the program, corresponding to a type definition, in which all forms of representation for the same kind of data are known.
\end{quote}

Unfortunately, Reynolds' paper is not widely known, even though it has been reprinted in two  collections of notable papers~\citep{reynol1978a,reynol1994}; many practitioners do not understand the fundamental distinction between type abstraction and procedural abstraction.
This distinction is important 
because there is a trade-off between the costs and benefits of the two approaches.
The benefit of procedural abstraction\,---\,the approach used by objects\,---\,is that it provides secure data abstraction without types, and allows multiple implementations of the same abstraction to co-exist; this supports change, and helps to keep software soft. 
The cost is that it is harder to program ``binary operations''\,---\,those that work on two or more objects\,---\,with maximal efficiency. 

William Cook made another attempt to explain this distinction in a 2009 OOPSLA Essay~\citep{cook2009}.
Cook coined the term ``autognostic'', meaning ``self-knowing'', for what Reynolds called ``decentralisation''.  
Autognosis means that an object can have detailed knowledge only of itself: all other objects are encapsulated.
Cook remarks: ``The converse is quite useful: any programming model that allows inspection of the representation of more than one abstraction at a time is not object-oriented.''

To see the benefits of procedural abstraction, consider
a selection of objects implementing numbers.  The objects $1$ and $2$ might include a representation as a machine integer, and methods \textsf{$+$},  \textsf{$-$}, \emph{etc}., that eventually use hardware arithmetic.  However, the object representing  $2^{67}$ might use a representation comprising three radix $2^{32}$ digits, and the methods for \textsf{$+$}  and \textsf{$-$}  would then perform multi-digit arithmetic.
This scheme, augmented with operations to convert machine integers to the multi-digit representation, is essentially how Smalltalk implements numbers.  
Notice that new representations of number objects can be added at any time, without having to change existing code, so long as the new objects provide the necessary methods.
Because the representation of an object can never be observed by any other object\footnote{To take advantage of the hardware arithmetic unit, that piece of hardware must naturally be allowed to ``see'' the representation of small integers.  Nevertheless, the representation of a object remains hidden from all other objects in the program.}, there is no need to protect it with an existential type.

The idea of procedural data abstraction was certainly in \Simula{} I\,---\,indeed, I believe that it was the genesis of \Simula{}~I.
As Dahl and Hoare observe in ``Hierarchical Program Structures''~\citep{dahl1972},
the key concept is already present in Algol 60: 
an Algol 60 block is a single mechanism that can contain both procedures and data.
The limitation of Algol is that blocks are constrained not to outlive their callers.
Dahl saw this, and ``set blocks free'' by devising a dynamic storage allocation scheme to replace Algol 60's stack~\citep{nygaar1981}.

However, procedural data structures were not the only kind of data aggregate supported by \Simula~67.  As I mentioned in Section~\ref{sec:SimulaSummary}, Dahl realised that \Simula's class construct could be used  to generate both records (unprotected, or protected by type abstraction) and objects (protected by procedural abstraction), and saw that as a strength of the mechanism; some later languages, notably C++, also support this dualism.
Records were made possible by \Simula~I's \textsf{inspect} statement, and later by \Simula~67's more flexible ``dot'' notation; these mechanisms were introduced specifically to expose the variables that would otherwise have been local to an object.
In contrast, Smalltalk made the instance variables of an object inaccessible to any other object.
If a Smalltalk object wanted to expose an instance variable, it could provide \emph{methods} that permitted the reading and writing of the variable, 
but the client would nevertheless see only methods, never variables.

The modern view of an object is that only the \emph{behaviour} of the object matters: whether that behaviour is ultimately supported by methods or variables is one of the hidden implementation details of the object.
The text-book illustration of this consists of two \code{Point} objects, one encapsulating a representation using cartesian coordinates, and the other encapsulating a representation using polar coordinates. 
Both objects can offer methods to retrieve, and to change, the \code{x}, \code{y}, \code{$\rho$} and \code{$\theta$} of the points.
Some of these methods will involve direct access to single instance variables, while other will involve trigonometric calculations and access to multiple instance variables.
Nevertheless, clients of the two objects will be able to treat them identically.

It seems to me that Dahl himself eventually came to believe that procedural abstraction was the fundamental contribution 
of the Simula languages.
In 2002, Dahl wrote~\citep{dahl2002}:
\begin{quote}
The most important new concept of Simula 67 is surely the idea of data structures with associated operators \ldots called objects. There is an important difference, except in trivial cases, between
\begin{itemize}
\item	\emph{the inside view} of an object, understood in terms of local variables, possibly initialising operations establishing an invariant, and implemented procedures operating on the variables maintaining the invariant, and
\item	 \emph{the outside view}, as presented by the remotely accessible procedures, including some generating mechanism, dealing with more ``abstract'' entities.
\end{itemize}
\end{quote}

There is one more consequence of the difference between type abstraction and procedural abstraction.
Type abstraction makes a fundamental distinction between the collection of representation variables\,---\,the value that is protected by the existential type\,---\,and the functions that operate on it.
Procedural abstraction makes no such distinction: it exposes \emph{only} methods, with the
consequence that there may \emph{be no representation variables at all!}

\begin{figure}[btp]
	\begin{center}
		\includegraphics[width=0.9\textwidth]{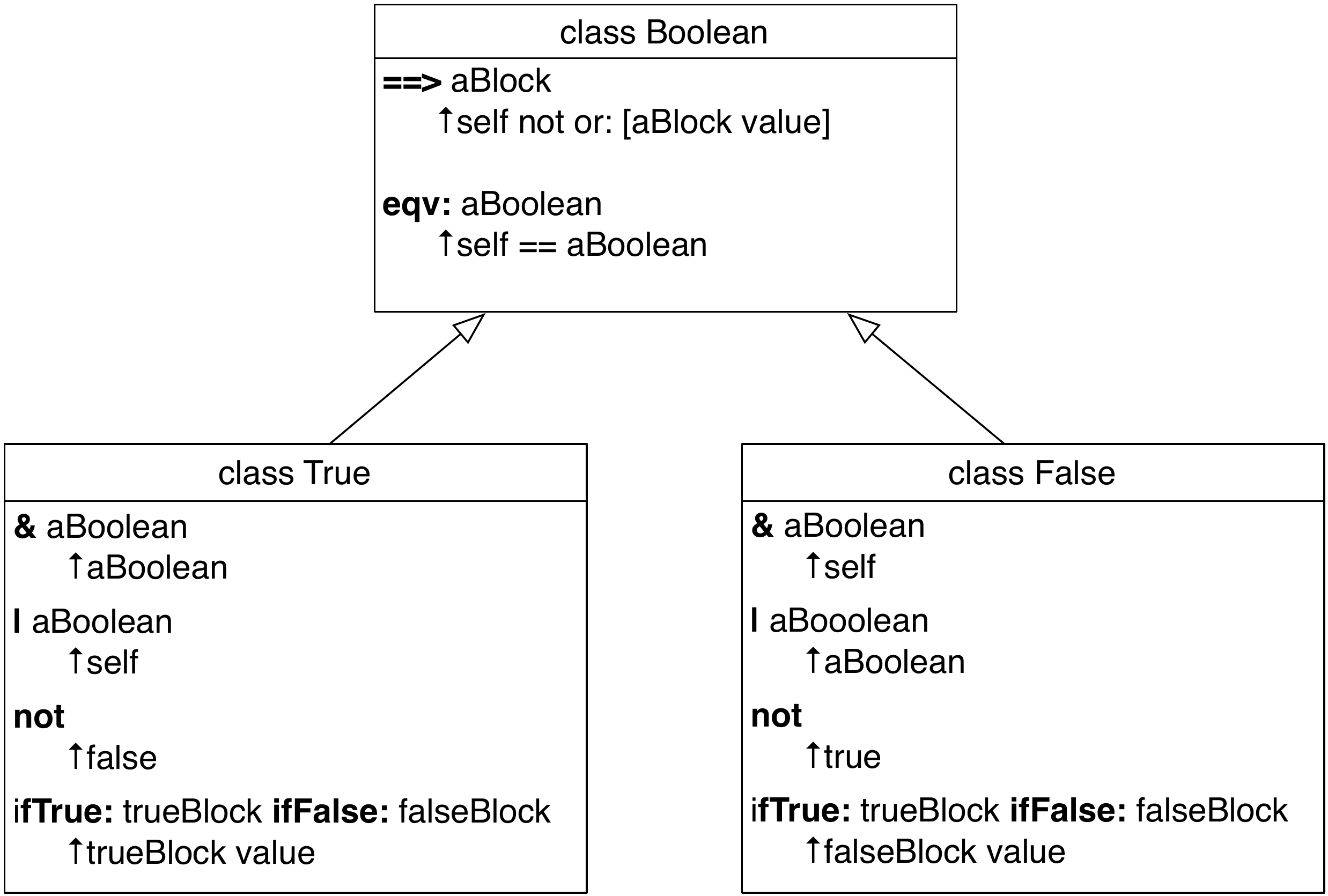}
	\end{center}
	\caption{An extract from the Boolean classes in Smalltalk.  \textsf{Boolean} is an abstract class with no instances; \textsf{true} and \textsf{false} are the only instances of the classes \textsf{True} and \textsf{False}, which inherit from \textsf{Boolean}.  There is no ``hidden state''; the methods themselves are sufficient to define the expected behaviour of the Booleans.}
	\label{fig:STBooleans}
\end{figure}

Perhaps the clearest example of this is the implementation of the Booleans in Smalltalk, which is outlined in Figure~\ref{fig:STBooleans}.
The objects \textsf{true} and \textsf{false} implement methods \textsf{\&}, \textsf{$|$}, \textsf{not}, \textsf{ifTrue:ifFalse:}, \textit{etc}.  (The colon in a
Smalltalk method name introduces an argument, in the same way that parentheses are used to indicate arguments in other languages, and $\uparrow$ means ``return from this method with the following answer''.)
For example, the \textsf{\&} method of \textsf{true} answers its argument, while the \textsf{$|$} method always answers \textsf{self}, \emph{i.e.}, \textsf{true}. 
The \textsf{ifTrue:ifFalse:} methods are a bit more complicated, because the arguments must be evaluated conditionally. The \textsf{ifTrue:ifFalse:} method of \textsf{false} ignores its first argument, evaluates its second (by sending it the message \textsf{value}), and returns the result, while the \textsf{ifTrue:ifFalse:} method of \textsf{true} does the reverse.

\section{Active Objects}
\noindent
Active objects seem to be one idea from \Simula{} that has been lost to the object-oriented community.
Activity was an important part of \Simula{}; after all, the original purpose of the language was to simulate activities from the real world.
\Simula's ``quasi-parallelism'' was a sweet-spot in 1961: it allowed programmers to think about concurrency while ignoring synchronisation.
Because another task could execute only when explicitly resumed, 
programmers could be confident that their data would not change ``out from under them'' at unexpected times.

Hewitt's Actor model~\citep{hewitt1973} built on this idea, as did Emerald~\citep{black1986c}, in which every object could potentially contain a process.
Other languages, like Erlang, have made the process the main focus of the language 
rather than the object~\citep{armstr2003}.
However, active objects have disappeared from ``mainstream'' object-oriented languages.
Since Smalltalk, processes and objects have been independent concepts.

Why are Smalltalk objects passive?
I don't know.
Perhaps Kay and Ingalls had a philosophical objection to combining what they saw as separate ideas.
Perhaps the realities of programming on the Alto set limits as to what was possible at that time.
Perhaps they wanted real processes, not coroutines; this requires explicit synchronisation, and thus a significantly different language design.
Smalltalk does indeed have processes, rather than coroutines.    
However, these processes are realised as separate objects, rather than being a part of every object.

\section{Classes and Objects}
\noindent
It's interesting to revisit Wegner's idea, discussed in Section~\ref{WegnerOO}, that   object-orientation is characterised by the existence of classes.
As we have seen, most of the concepts that we recognise as making up object-orientation came from the class concept of \Simula{}.
Nevertheless, as both Dahl and Nygaard would be quick to point out,
it's the dynamic objects, not the classes, that form the system model, and modelling was the \emph{raison d'\^{e}tre} of \Simula. 
Classes were interesting only as a way of creating the dynamic system model, which comprised interacting objects.
Classes are indeed a most convenient tool if you want hundreds of similar objects. But what if you want just one object?
In that case they impose a conceptual overhead.
The fact is that classes are meta, and meta isn't always better!
\label{sec:meta}

Classes are meta, relative to objects, because a class describes the implementation and behaviour of those objects that we call instances of the class, as well as providing a factory (procedure or method) that realises that description by actually making an instance.
If you need just one or two objects, it's simpler and clearer to describe those objects in the program directly,  rather than to describe a factory (that is, a class) to make them, and then use it once or twice.
This is the idea behind Self, Emerald, NewtonScript and JavaScript, which take the object, rather than the class, as their fundamental concept.
Which is not to say that classes are not useful: just that, at least in their r\^{o}le of object factories, they can be represented as objects.

In class-based languages, classes typically play many r\^{o}les.  Borning, in an early paper contrasting class-based and object-based languages~\citep{bornin1986}, lists eight r\^{o}les for Smalltalk classes.
The primary r\^{o}le of a class is acting as a factory that creates new objects; some of the other r\^{o}les served by Smalltalk classes include describing the representation of those objects, defining their message protocol, serving as repositories for the methods that implement that protocol, providing a taxonomy for classifying objects, and serving as the unit of inheritance.
Nevertheless, as we saw when discussing Ingalls' view of classification (Section~\ref{IngallsClassification} above), the taxonomy provided by classes, being based on implementation rather
than behaviour, is fundamentally at odds with the autognosis that other researchers, such as Cook, believe to be fundamental to object-orientation.

If the only thing that we can know about an object is its behaviour, the most useful way of classifying objects must be according to their behaviour.
This brings us back to types.
Recall that types are not needed to provide abstraction in object-oriented systems.
Nor are they needed to provide what Hoare called ``security'': that comes from procedural encapsulation.
The true r\^{o}le of types is to provide a \emph{behavioural} taxonomy: types provide a way of classifying objects so that programmers can better understand what r\^{o}les a parameter is expected to play, or what can be requested of an object returned as an answer from some ``foreign'' code.

Why is such a taxonomy useful? 
To add redundancy. 
A type annotation is an assertion about the value of a variable, no different in concept from asserting that a collection is empty or that an object is not \textsf{nil}.
Redundancy is in general a good thing:
it provides additional information for readers, and it means that more errors\,---\,at least those errors that manifest themselves as an inconsistency between the type annotation and the code\,---\,can be detected sooner.
Nevertheless, I am about to argue that types can be harmful. 

On the surface, this is a ridiculous argument.
If types add redundancy, and redundancy is good, how can types be harmful?
The problem is not that adding type annotations will mess up your program, but that adding types to a language can, unless one is very careful, mess up the \emph{language design}.

I have been an advocate of types for many years.
The Emerald type system~\citep{black1987,black1991c} was one of the ways in which the Emerald project attempted to improve on Smalltalk. 
Our original perception was that types would enable us to generate more efficient code.
However, as we implemented the language, we realised that behavioural types, that is, types that constrain the interface of an object, but not its implementation, did not help us to generate more efficient code.  
Emerald was efficient, but that was because we inferred implementation information, for example, that no reference to a particular object ever escaped an enclosing object.%
\footnote{In one place we did get efficiency from types, but that was because we cheated.
Emerald made it impossible to re-implement integers: a user-defined object would \emph{never} be recognised as having type \code{Integer}, even though it provided all of the necessary operations.  
This was cheating because Emerald types are supposed to constrain interface, not implementation.  The consequence was 
that we could indeed infer representation information from the \code{Integer} type,
and directly invoke machine operations to perform arithmetic.  
}
I spent many years working on type-checking and type systems, but have only recently really understood the sage advice that I was given by Butler Lampson in the late 1980s: stay away from types\,---\,they are just too difficult.

The difficulty springs from two sources.  One is G\"{o}del's second incompleteness theorem, which tells us that there \emph{will be} true facts about any (sufficiently rich) formal system that cannot be proved, no matter what system of proof we adopt.
In this case, the formal system is our programming language, and the system of proof is that language's type system.
The other source is 
our very natural desire to \emph{know} that our program won't go wrong, or at least that it won't go wrong in certain circumscribed ways.
This has led most statically-typed languages to adopt an interpretation of type-checking that I am going to refer to as ``The Wilson interpretation''.
The name honours Harold Wilson, prime minister of the UK 1964--70 and 1974--76.
Wilson fostered what has been called ``the Nanny State'', a social and political apparatus that said the government will look after you:  if it is even remotely possible that something will go wrong, we won't even let you try.
The Wilson interpretation of type-checking embraces two tenets:
first, that the type system must be complete, that is, every type assertion that can be made in it must be provably true or false, and second, that every part of the program must be type-checked before it can be executed.
The consequence is that if it is even remotely possible that something may go wrong when executing your program, then the language implementation won't even let you try to run it\,---\,assuming that the ``something'' that might go wrong is a thing over which the type system believes it has control.

The Wilson interpretation is not the only interpretation of type-checking.
At the other extreme is what I will call the ``Bush interpretation'',
honouring George W. Bush, president of the USA 2001--2009, who abolished many government regulations and weakened the enforcement of those that remained. 
The Bush interpretation of type checking is that type information is advisory.
The program should be allowed to do what you, the programmer, said that it should do.
The type-checker won't stop it; if you mess up, the PDIC\,---\,the Program Debugger and Interactive Checker\,---\,will bail you out.
The Bush interpretation amounts to not having static type-checking at all: the programmer is allowed to write type annotations in the program, but they won't be checked until runtime.

There is a third interpretation of type-checking that lies between these extremes.
This might be called the ``Proceed with caution'' approach.
If the type-checker has been unable to prove that there are no type errors in your program, you are given a warning. ``It may work; it may give you a run-time error. Good night, and good luck.''
I'm going to name this interpretation in honour of  Edward R. Murrow, 1908--1965, an American broadcaster, whose signature line that was.  
Like Murrow, I intend that phrase to be taken positively.

I'm referring to Wilson, Murrow, and Bush as \emph{interpretations} of type-checking, because I'm taking the view that the semantics of the language, in the absence of errors, is independent of its type system. 
In other words, the language has an untyped semantics, and
the r\^{o}le of the type system is to control how and when errors are reported, not to determine the meaning of the code.
                      There is an alternative view in which the meaning of a program depends on the types of its terms; programs that don't type-check \emph{have no semantics}.
In this view, Wilson, Murrow, and Bush define different languages, connected by a superset relation.
Since, as we have seen, types are not necessary to define object-based abstraction, 
I feel that the untyped view is simpler and more appropriate, but the comparison of these views is a topic worthy of its own essay, and I'm not going to discuss it further here.

So, under all three interpretations, an error-free program has the same meaning.
Under Wilson, we have conventional static typing:
an erroneous program will result in a static error, and won't be permitted to run.
Because of G\"odel's theorem, this means that \emph{some} error-free programs won't be permitted to run either, no matter how complex the type system becomes.

Under Bush, we have conventional dynamic typing: all checks will be performed at runtime\,---\,even those that are guaranteed to fail.
Enthusiasts for dynamic typing point out that
a counter-example is often more useful than a type-error message.
The disadvantage is that the programmer will not receive any compile-time warnings.
Under the Murrow interpretation, the programmer will get a mix of compile-time warnings and run-time checks.
A Murrow warning may say that a certain construct is provably a type error, or it may merely say that the type-checker has been unable to prove that it is type safe.
The difference between these two warnings is dramatic, but Wilson treats them as if they were the same, conflating a deficiency in the type system with a programming error.

Let me say here that I'm for Murrow!
I believe that the Murrow interpretation of types is the most useful, not only for programmers, but also for language designers.
Wilson's ``Nanny Statism'' is an invitation to mess up your language design.
The temptation for the language designer is to exclude any construct that can't be statically checked, regardless of how useful it may be to the practicing programmer.

I believe that \Simula{} was for Murrow too.
Recall that, according to Nygaard, the 
core ideas of SIMULA
were first modelling, and second, security.
Modelling meant that the actions and interactions of the objects that the program created represented the actions and interactions of the corresponding real-world objects.
Security meant that the behaviour of a program could be understood and explained entirely in terms of the semantics of the programming language in which it is written.
Modelling came first!
\Simula{} did not compromise its modelling ability to achieve security;
it compromised its run-time performance, by 
incorporating explicit checks when a construct necessary for modelling was not statically provable to be safe.

I feel that many current language designers are suffering from a Wilson obsession.
This obsession has resulted in
type systems of overwhelming complexity, and
languages that are larger, less regular, and less expressive than they need to be.
The fallacy is the assumption that, because type checking is good, \emph{more} type-checking is necessarily better.

Let's consider an example of how the insistence on static type-checking can mess up a language design.
I'm currently engaged, with Kim Bruce, James Noble and their students, in the design of a language for teaching programming.  
The language is called Grace, both to honour Rear Admiral Grace Hopper, and because we hope that it will enable the \emph{Graceful} expression of programs.
Grace is an object-oriented language, and contains an inheritance facility, inspired by early proposals by Antero Taivalsaari~\citep{taival1995}.
Here is an example:

\needspace{4\baselineskip}

\begin{codep}
class aDictionary.ofSize(initialSize) {
	inherits aHashtable.ofSize(initialSize) 
	method findIndex (predicate) is override { ... } 
	method at (key) put (value) is public { ... } 
	...
}
\end{codep}

In Grace, classes are nothing more than objects that have a single method that, when executed,
creates a new object.
This example declares a class object named \code{aDictionary}, which has an object-creation method with the name \code{ofSize}.  
Executing \textsf{aDictionary.ofSize(10)} will create and return a new object that
inherits (that is, contains copies of) all of the methods and instance variables of the object \textsf{aHashtable.ofSize(10)}, except that the given method for \textsf{findIndex()} will override that inherited from the Hashtable object, and the given method for \textsf{at()put()} will be added.

Checks that one might expect to take place on such a definition would include 
that \textsf{aHashtable} actually does have a method \textsf{ofSize}, and that the object answered by that method does have a \textsf{findIndex()} method and does \emph{not} have an \textsf{at()put()} method.  
There is no difficulty constructing a static type system to perform such checks so long as \textsf{aHashtable} is a globally known class.
However, suppose that I want to let the client choose the actual object that is extended, 
perhaps because prefix trees are better than hash tables for some applications.
In other words, suppose that I want to make the super-object a parameter:
\needspace{4\baselineskip}
\begin{codep}
class aDictionary.basedOn(superObj) { 
	inherits superObj
	method findIndex (predicate) is override { ... } 
	method at (key) put (value) is public { ... } 
	...
}
\end{codep}

Although the same implementation mechanisms will still work, constructing a static type system to perform the checking is no longer simple.  
What arguments may be substituted for \textsf{superObj}?
The requirement that \textsf{superObj} have a (possibly protected) method \textsf{findIndex}
(because of the \code{override} annotation) and that it \emph{not} have a method \textsf{at()put()} (because of the absence of an \code{override} annotation) cannot be captured by the usual notion of type, which holds only the information necessary to \emph{use} an object, not the information necessary to inherit from it.
One solution to the problem of type-checking this definition in a modular way is to invent a new notion of ``heir types'', that is, types that capture the information needed to check inheritance, such as the presence of a method that is either public or protected, or the absence of a method.

Another possible solution is to make classes a new sort of entity, different from objects, and to give them their own, non-Turing-complete sublanguage, including  class-specific function, parameter and choice mechanisms.  
A third possibility is to ban parametric super-objects.
How then does one use inheritance, that is, how does the programmer gain access to a superclass to extend?  The classical solution is to refer to classes by global variables, and to require that the superclass be globally known.
This is a premature commitment that runs in opposition to the late-binding that otherwise characterises object-orientation; its effect is to reduce reusability.
Some languages alleviate this problem by introducing yet another feature, known as open classes, which allows the programmer to add methods to, or override methods in, an existing globally-known class.  

Virtual classes, as found in \textsc{beta}~\citep{madsen1993}, are yet another approach to the problem of parametrizing inheritance.
A virtual class can be overridden in a subclass by a ``more specialised'' class.
Virtual classes don't by themselves \emph{solve} our problem: they need to
be accompanied by a set of type rules that specify when it is legal to override a virtual class.
The difficulty is that one set of rules\,---\,conventional subtyping\,---\,is appropriate for clients of the class, and another set of rules\,---\,heir typing\,---\,is appropriate for inheritors.
\textsc{beta} chooses to use co-variant method subtyping for virtual classes, that is, to allow the arguments of methods to be specialised along with their results.
This means that virtual classes cannot be guaranteed to be safe by a modular static analysis. Nevertheless, virtual classes are useful for modelling real systems;
\textsc{beta} follows \Simula{} in ranking modelling above static checking. 

Regardless of the path chosen, the resulting language is both larger and less-expressive than the simple parametric scheme sketched above.
Indeed, in the scheme suggested by my original example, a new object could extend \emph{any} existing object, and classes did not need to be ``special'' in any way; they could be ordinary objects that did not require any dedicated mechanisms for their creation, categorisation or use. 

Another example of this sort of problem is collections that are parametrized by types.  
If types are desirable, it certainly seems reasonable to give programmers the opportunity to parameterize their collection objects, so that a library implementor can offer \textsf{Bags} of \textsf{Employees} as well as \textsf{Bags} of \textsf{Numbers}.
The obvious solution is to represent types as objects, and use the normal method  and parameter mechanisms to parameterize collections\,---\,which is exactly what we did in Emerald~\citep{black2007}.
Unfortunately, the consequence of this is that type checking is no longer decidable~\citep{meyer1986}.
The reaction of the Wilsonian faction to this is to recoil in shock and horror.
After all, if the mission of types is to prevent the execution of any program that might possibly go wrong, and the very act of deciding whether the program will go wrong might not terminate, what is the language implementor to do?

A language that takes the Murrow interpretation of type can react to the possibility that  type-checking is statically undecidable more pragmatically.
The language implementation will inevitably use run-time type checks, to deal with situations in which the program cannot be  guaranteed to be free of type errors statically.
If a type assertion cannot be proved true or false after a reasonable amount of effort, it suffices to insert a run-time check.

An alternative way of parameterising objects by types is to invent a new parameter passing mechanism for types, with new syntax and semantics, but with restricted expressivity to ensure decidability.
Nevertheless, because of G\"odel's second incompleteness theorem, some programs will still be un-typeable.  The consequence is that the language becomes larger, while at the same time less expressive.

\section{The Future of Objects}
\noindent
Now I'm going to turn to some speculations about the programming languages of the future.
I could follow Feynman, and predict, with a certain confidence, that in 1000 years object-oriented programming will no longer exist as we know it.
This is either because our civilisation will have collapsed, or because it has not;
in the latter case humans will no longer be engaged in any activity resembling programming, which will instead be something that computers do for themselves.

I think that it is probably more useful, and certainly more within my capabilities, to look 10 or 20 years ahead.  
The changes in computing that will challenge our ability to write programs over that time span are already clear: the trend from multicore towards manycore, the need to control energy consumption,
the growth of mobility and  ``computing in the cloud'',  the demand for increased reliability and failure resilience, and the emergence of distributed software development teams.

\subsection{Multicore and Manycore}
\noindent
Although the exact number of ``cores'', that is, processing units, that will be present on the computer chips of 2021--31 is unknown, we can predict that with current technology trends we will be able to provide at least thousands, and perhaps hundreds of thousands.
It also seems clear that manycore chips will be much more heterogeneous than the two- and four-processor multicore chips of today, exactly because not all algorithms can take advantage of thousands of small, slow but power-efficient cores, and will need larger, more power-hungry cores to perform acceptably.
The degree of parallelism that will be available on commodity chips will thus depend both on the imperatives of electronic design and on whether we can solve the problem of writing programs that use many small cores effectively.

What do objects have to offer us in the era of manycore?
The answer seems obvious.
Processes that interact through messages, which are the essence of \Simula's view of computation, are exactly what populate a manycore chip.
The similarity becomes even clearer if we recall Kay's vision of objects as little computers.
Objects also offer us heterogeneity.
There are many different classes of objects interacting in a typical program, and there is no reason that these different classes of objects need be coded in the same programming language, or compiled to the same instruction set.  
This is because the ``encapsulation boundary'' around each object means that no other object need know about its implementation.
For example, objects that perform matrix arithmetic to calculate an image might be compiled to run on a general-purpose graphical processing unit, while other parts of the same application might be compiled to more conventional architectures.

\subsection{Controlling Energy Consumption}
\noindent
How does one write an energy-efficient program?  How, indeed, can one compare two programs for efficiency?  A common approach is to use a \emph{cost model}: a way of thinking about computation that is closely-enough aligned with the real costs of execution that it can help us to reason about performance.  Objects can provide us with a cost model for manycore computing.

Most current computing models date from the 1950s and treat computation as the expensive resource\,---\,which was the case when those models were developed, and computation made use of expensive and power-hungry valve or discrete-transistor logic.
In contrast, data movement is regarded as free, since it was implemented by copper wires, which were inexpensive compared to valves or transistors, and consumed little energy.
As a consequence, these models lead us to think, for example, of moving an operation out of a loop and caching the result in memory as an ``optimisation''.
The reality today is very different: computation is essentially free, because it happens ``in the cracks'' between data fetch and data store; on a clocked processor, there is little difference in energy consumption between performing an arithmetic operation and performing a no-op. 
In contrast, data movement is expensive: it takes both time and energy to propagate signals along wires.
So controlling the cost of computation on a manycore chip requires one to exercise control of data movement.
Today's programming languages are unable to even express the thing that needs to be carefully controlled: data movement.

I would like to propose an object-oriented cost model for manycore computing.
To \Simula's concurrent object model we need add only three things.
The first addition is to treat objects as enjoying true parallelism, rather than \Simula's quasi-parallelism.
The second and third additions are to attribute to each object a size and a spatial location.
Size is important because an object needs to fit into the cache memory available at its location.  
``Size'' includes not just the data inside the object, but also the code that makes up its method suite, since both must be cached.

With this model, local operations on an object can be regarded as free, since they require only local computations on local data.  
Optimising an object means reducing its size until it fits into the available cache, or, if this is not possible, partitioning the object into two or more smaller objects.
In contrast to local operations, requesting that a method be evaluated in another object may be costly, since this may require communicating with an object at a remote location.
The cost of such a message is proportional to the product of the amount of data in it and the distance to the receiver; we can imagine measuring cost in byte-nanometers.
The cost of a whole computation can be estimated as proportional to the number of messages sent and the cost of each message and reply.
We can reduce this cost by relocating objects so that frequent communication partners are close together, and by recomputing answers locally rather than requesting them from a remote location. 
Such a cost model does not, of course, capture all of the costs of a real computation, but it seems to me that it will provide a better guide for optimisation than models from the 1950s.

\subsection{Mobility and the Cloud}
\noindent
The world of mobile computing is a world in which communication is transient, failure is common, and replication and caching are the main techniques used to improve performance and availability.
The best models for accessing objects in such an environment seem to me to be those used for distributed version control, as realised in systems like subversion and git~\citep{mason2005,duan2010}.
In such systems, all objects are immutable; updating an object creates a new version.
Objects can be referred to in two ways: by version identifier and by name.
A version identifier is typically a long numeric string and refers to (a copy of) a particular immutable object.
In contrast, a name is designed for human consumption, but the binding of a name to a version object changes over time: when a name is resolved, it is by default bound to a recent version of the object that it describes.

Erlang uses a similar naming mechanism to realise its fail-over facility.
Erlang messages can be sent either to a process identifier or to a process name.
A particular process identifier will always refer to the same process; sending a message to a process identifier will fail if the process is no longer running. 
In contrast, a process name indirects through the name server, so messages sent to a process name will be delivered to the most recent process to register that name.
When a process crashes and its monitor creates a replacement process, 
the replacement will usually register itself with the name sever under the same name that was used by the crashed process.
This ensures that messages sent to the process name will still be delivered, although successive messages may be delivered to different processes~\citep{armstr2007}.
Perhaps a similar mechanism should be part of every object-oriented system?
This would mean that 
we would be able to reference an object either by a descriptor, such as ``Most recent version of the Oslo talk'', or by a unique identifier,  such as Object16x45d023f.
The former would resolve to a replica of a recent version; the latter would resolve to a specific object.

\subsection{Reliability}
\noindent
Total failures are easy for the programmer to handle, because there is nothing that the programmer can do!
However, they can be disastrous for the users of a computer system, who are prevented from doing their work. 
As distributed systems have become ubiquitous, total failures have become rare, and partial failures much more common.
This trend will accelerate as the size of the features on a chip decreases, and massively manycore chips proliferate. 
The reliability of individual transistors on such chips will be much lower than we are accustomed to today: this is a consequence of the physics of doping.  By 2014, if there are 100 billion transistors on a chip, 20 percent of them will not work when it is fabricated, and another 10 percent will fail over the life of the chip\footnote{These predictions are from Shekhar Borkar, an Intel fellow and Director of the Microprocessor Technology lab.  They were presented by Bob Colwell at FCRC, San Diego, 2007.}.  
This means two things:  that processor design will have to change quite radically, and that single chips will have reliability characteristics similar to today's Internet, where computers may fail at any time. 

What techniques are available for dealing with partial failures?  
It is often possible to mask a failure using replication in time or space, but this is not always wise.  
Resending lost messages is not a good idea if they contain out-of date information; it may be better to  send a new message containing fresh information.
Similarly, it may be more cost-effective to recompute lost data than to retrieve it from a remote replica.
So automatic masking of failure by the programming model is not a good idea.
Instead, it seems to me that a variety of facilities for dealing with failures must be visible in the programming model, so that programmers can choose whether or not to use them.
What, then, should be the unit of failure in an object-oriented model of computation?
Is it the object, or is there some other, larger, unit? 
Whatever unit we choose, it must ``leak'' failure, in the sense that its clients must be able to see it fail, and then take some action appropriate to the context.

\subsection{Distributed Development Teams}
\noindent
With the ubiquity of computing equipment and the globalisation of industry, distributed software development teams have become common.  You may wonder what this trend has to do with objects.
The answer is packaging: 
collaborating in loosely-knit teams demands better tools for packaging code and sharing it between the parts of a distributed team.
Modules, by which I mean the unit of code sharing, are typically parametrized by other modules.
To take full advantage of the power of objects, 
module parameters should be bound at the time that a module is used, not when it is written.
In other words, module parameters should be ``late bound''.
\hyphenation{name-space}%
This means that there is no need for a global namespace in the programming language itself; URLs
or versioned objects provide a perfectly adequate mechanism for referring to the arguments of modules.

\section{The Value of Dynamism}
\noindent
Before I conclude, I'm going to offer some speculations on the value of dynamism.
These speculations are motivated by the fact that in designing \Simula,
Dahl recognised that the \emph{runtime structures} of Algol 60 already contained the mechanisms that were necessary for simulation, and by Nygaard and Dahl's understanding that it is the \emph{runtime behaviour} of a simulation program that models the real world, not the program's text.
I see a similar recognition of the value of dynamism in 
Agile software development, a methodology in which a program is developed in small increments, in close consultation with the customer.
The idea is that a primitive version of the code runs at the end of the first week, and new functionality is added every week, under the guidance of the customer, who determines which functions have the greatest value.
Extensive test suites make sure that the old functionality continues to work while new functionality is added.

If you have never tried such a methodology, you may wonder how it could possibly work!
After all, isn't it important to \emph{design} the program?
The answer is yes: design is important.
In fact, design is \emph{so} important, that Agile practitioners don't do it only at the start of the software creation process, when they know nothing about the program. 
Instead, they design every day: the program is continuously re-designed as the programmers learn from the code, and from the behaviour of the running system.

What I've learned from watching this process is that the program's run-time behaviour is a powerful teaching tool.
This is obvious if you view programs as existing to control computers, but perhaps less obvious if you view programs as static descriptions of a system, like a set of mathematical equations.
As Dahl and Nygaard discovered, it is a program's behaviour, not the program itself, that models the real-world system.
The program's text is a meta-description of the program behaviour, and it is not always easy to infer the behaviour from the meta-description.

As a teacher of object-oriented programming, 
I know that I have succeeded when students anthropomorphise their objects, that is, when they turn to their partners and speak of one object asking another object to do something.
I have found that this happens more often, and more quickly, when I teach with Smalltalk than when I teach with Java: 
Smalltalk programmers tend to talk about objects, while Java programmers tend to talk about classes.
I suspect that this is because Smalltalk is the more dynamic language:
the language and the programming environment are designed to help programmers interact with objects, as well as with code.
Indeed, I am tempted to define a ``Dynamic Programming Language'' as one designed to help the programmer learn from the run-time behaviour of the program.

\section{Summary}
\noindent
Fifty years ago, Dahl and Nygaard had some profound insights about the nature of both computation and human understanding.
``Modelling the world'' is not only a powerful technique for designing and re-designing a computer program: it is also one of the most effective ways yet found of communicating that design to other humans.

What are the major concepts of object-orientation?
The answer depends on the social and political context.
Dahl listed five, including the use of objects as record structures and as procedurally-encapsulated data abstractions, and the use of inheritance for incremental definition.
Some of what he though were key ideas, such as active objects and objects as modules, have been neglected over the last 30 years, but may yet be revived as the context in which we program becomes increasingly distributed and heterogeneous, in terms of both execution platform and programming team.
It certainly seems to me that, after 50 years, there are still ideas in \Simula{} that can be mined to solve twenty-first century problems.
There may not be any programming a thousand years from now,
but I'm willing to wager that some form of Dahl's ideas will still be familiar to programmers in fifty years, when \Simula{} celebrates its centenary.

\section*{Acknowledgements}
\noindent
I thank Stein Krogdahl, Olaf Owe and the committee of  FCT11 for honouring me with the invitation to speak at the scientific opening of the Ole-Johan Dahl hus, as well as supplying me with information about Dahl and \Simula.  I also thank Olaf for convincing me to turn my lecture into this article.
William Cook was an encouraging and through critic; 
Ivan Sutherland helped me understand where the costs lie in modern processors, and provided useful feedback on a draft of my manuscript.
David Ungar shared many insights with me and helped to improve both my lecture and this article.  The anonymous referees were especially helpful in pointing out errors of fact and 
passages in need of clarification.





\bibliographystyle{model1-num-names}
\bibliography{Bibliography/Omnibus}

\begin{thebibliography}{46}
\expandafter\ifx\csname natexlab\endcsname\relax\def\natexlab#1{#1}\fi
\providecommand{\bibinfo}[2]{#2}
\ifx\xfnm\relax \def\xfnm[#1]{\unskip,\space#1}\fi
\bibitem[{{FCT 2011}(2011)}]{fct2011}
\bibinfo{author}{{FCT 2011}}, \bibinfo{title}{Scientific opening of the
  {Ole-Johan Dahl} building}, \bibinfo{year}{2011}. \bibinfo{note}{Web page
  last visited 2 March 2013.
  \url{http://fct11.ifi.uio.no/index.php?n=General.SocialEvents}}.
\bibitem[{Nygaard and Dahl(1981)}]{nygaar1981}
\bibinfo{author}{K.~Nygaard}, \bibinfo{author}{O.-J. Dahl},
\newblock \bibinfo{title}{The development of the {SIMULA} languages},
\newblock in: \bibinfo{editor}{R.~L. Wexelblat} (Ed.),
  \bibinfo{booktitle}{History of programming languages I},
  \bibinfo{publisher}{ACM}, \bibinfo{address}{New York, NY, USA},
  \bibinfo{year}{1981}, pp. \bibinfo{pages}{439--480}.
\bibitem[{{University of Oslo, Department of Informatics}(2011)}]{UnOslo2011}
\bibinfo{author}{{University of Oslo, Department of Informatics}},
  \bibinfo{title}{{Kristen Nygaard\,---\,Education and Career}},
  \bibinfo{year}{2011}. \bibinfo{note}{Web page last visited 25 December 2011.
  \url{http://www.mn.uio.no/ifi/english/about/kristen-nygaard/career/}}.
\bibitem[{Feynman(1961)}]{feynma1961}
\bibinfo{author}{R.~Feynman}, \bibinfo{title}{Speech at {Centenial}
  {Celebration} of {Massachusetts Institute of Technology} (version {C})},
  \bibinfo{year}{1961}. \bibinfo{note}{The Feynman Archives at California
  Institute of Technology}.
\bibitem[{Hoare(1973)}]{HoareHints}
\bibinfo{author}{C.~Hoare}, \bibinfo{title}{Hints on Programming Language
  Design}, \bibinfo{type}{Memo} \bibinfo{number}{{AIM}-224}, Stanford
  Artificial Intelligence Laboratory, \bibinfo{year}{1973}.
  \bibinfo{note}{Invited address, 1st POPL conference}.
\bibitem[{Radin(1978)}]{radin1978}
\bibinfo{author}{G.~Radin},
\newblock \bibinfo{title}{The early history and characteristics of {PL/I}},
\newblock \bibinfo{journal}{SIGPLAN Notices} \bibinfo{volume}{13}
  (\bibinfo{year}{1978}) \bibinfo{pages}{227--241}.
\bibitem[{Dahl and Nygaard(1967)}]{dahl1967}
\bibinfo{author}{O.-J. Dahl}, \bibinfo{author}{K.~Nygaard},
\newblock \bibinfo{title}{Class and subclass declarations},
\newblock in: \bibinfo{editor}{J.~N. Buxton} (Ed.),
  \bibinfo{booktitle}{Simulation Programming Languages, Proceedings from the
  IFIP working conference in Oslo}, \bibinfo{publisher}{North Holland},
  \bibinfo{address}{Amsterdam}, \bibinfo{year}{1967}, pp.
  \bibinfo{pages}{158--174}.
\bibitem[{Dahl(1981)}]{dahl1981}
\bibinfo{author}{O.-J. Dahl},
\newblock \bibinfo{title}{Transcript of discussant's remarks},
\newblock in: \bibinfo{editor}{R.~L. Wexelblat} (Ed.),
  \bibinfo{booktitle}{History of programming languages I},
  \bibinfo{publisher}{ACM}, \bibinfo{address}{New York, NY, USA},
  \bibinfo{year}{1981}, pp. \bibinfo{pages}{488--490}.
\bibitem[{Bracha et~al.(2010)Bracha, von~der Ah{\'e}, Bykov, Kashai, Maddox,
  and Miranda}]{bracha2010}
\bibinfo{author}{G.~Bracha}, \bibinfo{author}{P.~von~der Ah{\'e}},
  \bibinfo{author}{V.~Bykov}, \bibinfo{author}{Y.~Kashai},
  \bibinfo{author}{W.~Maddox}, \bibinfo{author}{E.~Miranda},
\newblock \bibinfo{title}{Modules as objects in newspeak},
\newblock in: \bibinfo{editor}{T.~D'Hondt} (Ed.), \bibinfo{booktitle}{ECOOP},
  volume \bibinfo{volume}{6183} of \textit{\bibinfo{series}{Lecture Notes in
  Computer Science}}, \bibinfo{publisher}{Springer}, \bibinfo{year}{2010}, pp.
  \bibinfo{pages}{405--428}.
\bibitem[{Black et~al.(2012)Black, Bruce, Homer, and Noble}]{black2012}
\bibinfo{author}{A.~P. Black}, \bibinfo{author}{K.~B. Bruce},
  \bibinfo{author}{M.~Homer}, \bibinfo{author}{J.~Noble},
\newblock \bibinfo{title}{{Grace}: the absence of (inessential) difficulty},
\newblock in: \bibinfo{booktitle}{Onward! '12: Proceedings 12th SIGPLAN Symp.
  on New Ideas in Programming and Reflections on Software},
  \bibinfo{publisher}{ACM}, \bibinfo{address}{New York, NY},
  \bibinfo{year}{2012}, pp. \bibinfo{pages}{85--98}.
\bibitem[{Dahl and Hoare(1972)}]{dahl1972}
\bibinfo{author}{O.-J. Dahl}, \bibinfo{author}{C.~Hoare},
\newblock \bibinfo{title}{Hierarchical program structures},
\newblock in: \bibinfo{booktitle}{Structured Programming},
  \bibinfo{publisher}{Academic Press}, \bibinfo{year}{1972}, pp.
  \bibinfo{pages}{175--220}.
\bibitem[{Hoare(1965)}]{hoare1965}
\bibinfo{author}{C.~A.~R. Hoare},
\newblock \bibinfo{title}{Record handling},
\newblock \bibinfo{journal}{ALGOL Bull.} \bibinfo{volume}{21}
  (\bibinfo{year}{1965}) \bibinfo{pages}{39--69}.
\bibitem[{Genuys(1968)}]{genuys1968}
\bibinfo{editor}{F.~Genuys} (Ed.), \bibinfo{title}{Programming Languages: NATO
  Advanced Study Institute}, \bibinfo{publisher}{Academic Press},
  \bibinfo{address}{London, New York}, \bibinfo{year}{1968}.
\bibitem[{Dahl(2002)}]{dahl2002}
\bibinfo{author}{O.-J. Dahl},
\newblock \bibinfo{title}{The roots of object-oriented programming: the
  {Simula} language},
\newblock in: \bibinfo{editor}{M.~Broy}, \bibinfo{editor}{E.~Denert} (Eds.),
  \bibinfo{booktitle}{Software Pioneers: Contributions to Software
  Engineering}, \bibinfo{publisher}{Springer-Verlag}, \bibinfo{address}{Berlin,
  Heidelberg}, \bibinfo{year}{2002}, pp. \bibinfo{pages}{79--90}.
\bibitem[{Wegner(1987)}]{wegner1987}
\bibinfo{author}{P.~Wegner},
\newblock \bibinfo{title}{Dimensions of object-based language design},
\newblock in: \bibinfo{editor}{N.~Meyrowitz} (Ed.),
  \bibinfo{booktitle}{Proceedings Second ACM Conference on Object-Oriented
  Programming Systems, Languages and Applications}, \bibinfo{publisher}{ACM
  Press}, \bibinfo{address}{Orlando, Florida}, \bibinfo{year}{1987}, pp.
  \bibinfo{pages}{168--182}.
\bibitem[{Black et~al.(1986)Black, Hutchinson, Jul, and Levy}]{black1986c}
\bibinfo{author}{A.~P. Black}, \bibinfo{author}{N.~Hutchinson},
  \bibinfo{author}{E.~Jul}, \bibinfo{author}{H.~Levy},
\newblock \bibinfo{title}{Object structure in the {Emerald} system},
\newblock in: \bibinfo{booktitle}{Proceedings First ACM Conference on
  Object-Oriented Programming Systems, Languages and Applications},
  \bibinfo{publisher}{ACM Press}, \bibinfo{address}{Portland, Oregon},
  \bibinfo{year}{1986}, pp. \bibinfo{pages}{78--86}.
\bibitem[{Black et~al.(1987)Black, Hutchinson, Jul, Levy, and
  Carter}]{black1987}
\bibinfo{author}{A.~P. Black}, \bibinfo{author}{N.~Hutchinson},
  \bibinfo{author}{E.~Jul}, \bibinfo{author}{H.~M. Levy},
  \bibinfo{author}{L.~Carter},
\newblock \bibinfo{title}{Distribution and abstract types in {Emerald}},
\newblock \bibinfo{journal}{IEEE Trans. Software Engineering}
  \bibinfo{volume}{SE-13} (\bibinfo{year}{1987}) \bibinfo{pages}{65--76}.
\bibitem[{Cook(1989)}]{Cook1989}
\bibinfo{author}{W.~R. Cook}, \bibinfo{title}{A Denotational Semantics of
  Inheritance}, Ph.D. thesis, Brown University, Department of Computer Science,
  \bibinfo{year}{1989}.
\bibitem[{Armstrong(2007)}]{armstr2007}
\bibinfo{author}{J.~Armstrong}, \bibinfo{title}{Programming {Erlang}: Software
  for a concurrent world}, \bibinfo{publisher}{Pragmatic Bookshelf},
  \bibinfo{year}{2007}.
\bibitem[{Archer(2011)}]{archer2011}
\bibinfo{author}{B.~Archer}, \bibinfo{title}{Programming quotations},
  \bibinfo{year}{2011}. \bibinfo{note}{Web page last visited 22 January 2012.
  \url{http://www.bobarcher.org/software/programming_quotes.html}}.
\bibitem[{Wadler(1992{\natexlab{a}})}]{wadler1992a}
\bibinfo{author}{P.~Wadler},
\newblock \bibinfo{title}{The essence of functional programming},
\newblock in: \bibinfo{booktitle}{Conference Record of the Nineteenth ACM
  Symposium on Principles of Programming Languages}, \bibinfo{publisher}{ACM
  Press}, \bibinfo{address}{Albuquerque, NM},
  \bibinfo{year}{1992}{\natexlab{a}}, pp. \bibinfo{pages}{1--14}.
\bibitem[{Wadler(1992{\natexlab{b}})}]{wadler1992}
\bibinfo{author}{P.~Wadler},
\newblock \bibinfo{title}{Comprehending monads},
\newblock \bibinfo{journal}{Mathematical Structures in Computer Science}
  \bibinfo{volume}{2} (\bibinfo{year}{1992}{\natexlab{b}})
  \bibinfo{pages}{461--493}. \bibinfo{note}{Originally published in ACM
  Conference on Lisp and Functional Programming, June 1990}.
\bibitem[{Kay(1993)}]{kay1993a}
\bibinfo{author}{A.~C. Kay},
\newblock \bibinfo{title}{The early history of {Smalltalk}},
\newblock in: \bibinfo{booktitle}{The second ACM SIGPLAN conference on History
  of programming languages}, HOPL-II, \bibinfo{publisher}{ACM Press},
  \bibinfo{address}{New York, NY, USA}, \bibinfo{year}{1993}, pp.
  \bibinfo{pages}{511--598}.
\bibitem[{Snyder(1991)}]{snyder1991a}
\bibinfo{author}{A.~Snyder}, \bibinfo{title}{The Essence of Objects: Common
  Concepts and Terminology}, \bibinfo{type}{Technical Report}
  \bibinfo{number}{HPL-91-50}, Hewlett Packard Laboratories,
  \bibinfo{year}{1991}.
\bibitem[{Snyder(1993)}]{snyder1993}
\bibinfo{author}{A.~Snyder},
\newblock \bibinfo{title}{The essence of objects: Concepts and terms},
\newblock \bibinfo{journal}{IEEE Software} \bibinfo{volume}{10}
  (\bibinfo{year}{1993}) \bibinfo{pages}{31--42}.
\bibitem[{Krogdahl(2010)}]{Krogda2010}
\bibinfo{author}{S.~Krogdahl}, \bibinfo{title}{Concepts and terminology in the
  {Simula} programming language\,---\,an introduction for new readers of
  {Simula} literature}, \bibinfo{year}{2010}.
  \bibinfo{note}{\url{http://folk.uio.no/simula67/Archive/concepts.pdf}}.
\bibitem[{Ingalls(1981)}]{ingall1981}
\bibinfo{author}{D.~H. Ingalls},
\newblock \bibinfo{title}{Design principles behind {Smalltalk}},
\newblock \bibinfo{journal}{Byte} \bibinfo{volume}{6} (\bibinfo{year}{1981})
  \bibinfo{pages}{286--298}.
\bibitem[{Borning and Ingalls(1982)}]{bornin1982}
\bibinfo{author}{A.~H. Borning}, \bibinfo{author}{D.~H.~H. Ingalls},
\newblock \bibinfo{title}{A type declaration and inference system for
  {Smalltalk}},
\newblock in: \bibinfo{booktitle}{Conference Record of the Ninth ACM Symposium
  on Principles of Programming Languages}, \bibinfo{publisher}{ACM Press},
  \bibinfo{address}{Albuquerque, NM, USA}, \bibinfo{year}{1982}, pp.
  \bibinfo{pages}{133--141}.
\bibitem[{Hoare(1972)}]{hoare1972}
\bibinfo{author}{C.~Hoare},
\newblock \bibinfo{title}{Proof of correctness of data representations},
\newblock \bibinfo{journal}{Acta Informatica} \bibinfo{volume}{1}
  (\bibinfo{year}{1972}) \bibinfo{pages}{271--281}.
\bibitem[{Parnas(1972)}]{parnas1972b}
\bibinfo{author}{D.~L. Parnas},
\newblock \bibinfo{title}{On the criteria to be used in decomposing systems
  into modules},
\newblock \bibinfo{journal}{Comm. ACM} \bibinfo{volume}{15}
  (\bibinfo{year}{1972}) \bibinfo{pages}{1053--1058}.
\bibitem[{Liskov et~al.(1977)Liskov, Snyder, Atkinson, and
  Schaffert}]{Liskov77}
\bibinfo{author}{B.~Liskov}, \bibinfo{author}{A.~Snyder},
  \bibinfo{author}{R.~Atkinson}, \bibinfo{author}{C.~Schaffert},
\newblock \bibinfo{title}{Abstraction mechanisms in {CLU}},
\newblock \bibinfo{journal}{Comm. ACM} \bibinfo{volume}{20}
  (\bibinfo{year}{1977}) \bibinfo{pages}{564--576}.
\bibitem[{Church(1941)}]{church1941}
\bibinfo{author}{A.~Church}, \bibinfo{title}{The Calculi of Lambda-Conversion},
  volume~\bibinfo{volume}{6} of \textit{\bibinfo{series}{Annals of Mathematical
  Studies}}, \bibinfo{publisher}{Princeton University Press},
  \bibinfo{year}{1941}.
\bibitem[{Reynolds(1975)}]{reynol1975}
\bibinfo{author}{J.~C. Reynolds},
\newblock \bibinfo{title}{User-defined types and procedural data structures as
  complementary approaches to data abstraction},
\newblock in: \bibinfo{editor}{S.~A. Schuman} (Ed.),
  \bibinfo{booktitle}{Conference on New Directions in Algorithmic Languages},
  \bibinfo{organization}{IFIP Working Group 2.1 on Algol},
  \bibinfo{publisher}{INRIA}, \bibinfo{address}{Munich, Germany},
  \bibinfo{year}{1975}, pp. \bibinfo{pages}{157--168}.
\bibitem[{Reynolds(1978)}]{reynol1978a}
\bibinfo{author}{J.~C. Reynolds},
\newblock \bibinfo{title}{User-defined types and procedural data structures as
  complementary approaches to data abstraction},
\newblock in: \bibinfo{editor}{D.~Gries} (Ed.), \bibinfo{booktitle}{Programming
  Methodology: A Collection of Articles by members of IFIP WG2.3}, Texts and
  Monographs in Computer Science, \bibinfo{publisher}{Springer Verlag},
  \bibinfo{address}{New York, Heidelberg, Berlin}, \bibinfo{year}{1978}, pp.
  \bibinfo{pages}{309--317}.
\bibitem[{Reynolds(1994)}]{reynol1994}
\bibinfo{author}{J.~C. Reynolds},
\newblock \bibinfo{title}{User-defined types and procedural data structures as
  complementary approaches to data abstraction},
\newblock in: \bibinfo{editor}{C.~A. Gunter}, \bibinfo{editor}{J.~C. Mitchell}
  (Eds.), \bibinfo{booktitle}{Theoretical Aspects of Object-Oriented
  Programming: Types, Semantics, and Language Design}, \bibinfo{publisher}{MIT
  Press}, \bibinfo{address}{Cambridge, MA}, \bibinfo{year}{1994}, pp.
  \bibinfo{pages}{13--23}.
\bibitem[{Cook(2009)}]{cook2009}
\bibinfo{author}{W.~R. Cook},
\newblock \bibinfo{title}{On understanding data abstraction, revisited},
\newblock in: \bibinfo{editor}{S.~Arora}, \bibinfo{editor}{G.~T. Leavens}
  (Eds.), \bibinfo{booktitle}{Proceedings 24th ACM Conference on
  Object-Oriented Programming, Systems, Languages, and Applications},
  \bibinfo{publisher}{ACM}, \bibinfo{year}{2009}, pp.
  \bibinfo{pages}{557--572}.
\bibitem[{Hewitt et~al.(1973)Hewitt, Bishop, and Steiger}]{hewitt1973}
\bibinfo{author}{C.~Hewitt}, \bibinfo{author}{P.~Bishop},
  \bibinfo{author}{R.~Steiger},
\newblock \bibinfo{title}{A universal modular actor formalism for artificial
  intelligence},
\newblock in: \bibinfo{editor}{N.~J. Nilsson} (Ed.),
  \bibinfo{booktitle}{IJCAI}, \bibinfo{publisher}{William Kaufmann},
  \bibinfo{address}{Stanford University, California}, \bibinfo{year}{1973}, pp.
  \bibinfo{pages}{235--245}.
\bibitem[{Armstrong(2003)}]{armstr2003}
\bibinfo{author}{J.~Armstrong},
\newblock \bibinfo{title}{Concurrency oriented programming in {Erlang}},
\newblock in: \bibinfo{booktitle}{Fr{\"u}hjahrsfachgespr{\"a}ch 2003 (FG
  2003)}, \bibinfo{publisher}{German Unix Users Group},
  \bibinfo{address}{K{\"o}ln, Germany}, \bibinfo{year}{2003}.
\bibitem[{Borning(1986)}]{bornin1986}
\bibinfo{author}{A.~Borning},
\newblock \bibinfo{title}{Classes versus prototypes in object-oriented
  languages},
\newblock in: \bibinfo{booktitle}{FJCC}, \bibinfo{publisher}{IEEE Computer
  Society}, \bibinfo{year}{1986}, pp. \bibinfo{pages}{36--40}.
\bibitem[{Black and Hutchinson(1991)}]{black1991c}
\bibinfo{author}{A.~P. Black}, \bibinfo{author}{N.~Hutchinson},
  \bibinfo{title}{Typechecking Polymorphism in Emerald},
  \bibinfo{type}{Technical Report} \bibinfo{number}{Technical Report CRL 91/1
  (Revised)}, Digital Equipment Corporation Cambridge Research Laboratory,
  \bibinfo{year}{1991}.
\bibitem[{Taivalsaari(1995)}]{taival1995}
\bibinfo{author}{A.~Taivalsaari},
\newblock \bibinfo{title}{Delegation versus concatenation, or cloning is
  inheritance too},
\newblock \bibinfo{journal}{SIGPLAN OOPS Mess.} \bibinfo{volume}{6}
  (\bibinfo{year}{1995}) \bibinfo{pages}{20--49}.
\bibitem[{Madsen(1993)}]{madsen1993}
\bibinfo{author}{O.~L. Madsen},
\newblock \bibinfo{title}{An overview of \textsc{BETA}},
\newblock in: \bibinfo{editor}{J.~Knudsen}, \bibinfo{editor}{O.~Madsen},
  \bibinfo{editor}{B.~Magnusson} (Eds.), \bibinfo{booktitle}{Object-Oriented
  Environments}, \bibinfo{publisher}{Prentice Hall},
  \bibinfo{address}{Englewood Cliffs, NJ}, \bibinfo{year}{1993}, pp.
  \bibinfo{pages}{99--118}.
\bibitem[{Black et~al.(2007)Black, Hutchinson, Jul, and Levy}]{black2007}
\bibinfo{author}{A.~P. Black}, \bibinfo{author}{N.~C. Hutchinson},
  \bibinfo{author}{E.~Jul}, \bibinfo{author}{H.~M. Levy},
\newblock \bibinfo{title}{The development of the {Emerald} programming
  language},
\newblock in: \bibinfo{editor}{B.~G. Ryder}, \bibinfo{editor}{B.~Hailpern}
  (Eds.), \bibinfo{booktitle}{HOPL III: Proceedings Third ACM SIGPLAN
  conference on History of Programming Languages}, \bibinfo{publisher}{ACM
  Press}, \bibinfo{address}{San Diego, CA}, \bibinfo{year}{2007}, pp.
  \bibinfo{pages}{11--1--11--51}.
\bibitem[{Meyer and Reinhold(1986)}]{meyer1986}
\bibinfo{author}{A.~R. Meyer}, \bibinfo{author}{M.~B. Reinhold},
\newblock \bibinfo{title}{`{Type}' is not a {Type}: Preliminary report},
\newblock in: \bibinfo{booktitle}{Conference Record of the Thirteenth ACM
  Symposium on Principles of Programming Languages}, \bibinfo{publisher}{ACM
  Press}, \bibinfo{address}{St Petersburg Beach, FL, USA},
  \bibinfo{year}{1986}, pp. \bibinfo{pages}{287--295}.
\bibitem[{Mason(2005)}]{mason2005}
\bibinfo{author}{M.~Mason}, \bibinfo{title}{Pragmatic Version Control Using
  Subversion}, Pragmatic Bookshelf, \bibinfo{publisher}{Pragmatic Programmers},
  \bibinfo{year}{2005}.
\bibitem[{Duan(2012)}]{duan2010}
\bibinfo{author}{C.~Duan}, \bibinfo{title}{Understanding {Git}: Repositories},
  \bibinfo{year}{2012}.
  \bibinfo{note}{\url{http://www.eecs.harvard.edu/~cduan/technical/git/git-1.s%
html}. Accessed on 3 Aug 2012}.

\end{thebibliography}







\end{document}